\documentclass[twocolumn]{aastex631}
\usepackage{comment}
\usepackage[encapsulated]{CJK}
\usepackage{ucs}
\usepackage[utf8]{inputenc}
\usepackage[T1]{fontenc}
\usepackage{graphicx}
\usepackage{float}
\usepackage{amsmath}
\usepackage{bm}
\usepackage{mathrsfs}
\usepackage{subfigure}
\usepackage{xspace}
\usepackage{nicefrac}

\graphicspath{{figures/}}

\newcommand{\cntext}[1]{\begin{CJK}{UTF8}{gbsn}#1\end{CJK}}

\defcitealias{PaperI}{EHTC~I}
\defcitealias{PaperII}{EHTC~II}
\defcitealias{PaperIII}{EHTC~III}
\defcitealias{PaperIV}{EHTC~IV}
\defcitealias{PaperV}{EHTC~V}
\defcitealias{PaperVI}{EHTC~VI}
\defcitealias{PaperVII}{EHTC~VII}
\defcitealias{PaperVIII}{EHTC~VIII}

\newcommand{\bhspin}{a_*}
\newcommand{\rhigh}{R_{\mathrm{high}}}
\newcommand{\rlow}{R_{\mathrm{low}}}
\newcommand{\bhname}{M87$^*$\xspace}
\newcommand{\sgra}{Sgr~A$^*$\xspace}

\newcommand{\ipole}{\texttt{ipole}\xspace}
\newcommand{\ipoleIL}{{\tt{ipole-IL}}\xspace}
\newcommand{\iharm}{{\tt{iharm3D}}\xspace}
\newcommand{\grtrans}{{\tt{grtrans}}\xspace}
\newcommand{\raptor}{{\tt{RAPTOR}}\xspace}
\newcommand{\odyssey}{{\tt{Odyssey}}\xspace}
\newcommand{\bhoss}{{\tt{BHOSS}}\xspace}

\newcommand{\mnet}{|m|_{\mathrm{net}}}
\newcommand{\vnet}{v_{\mathrm{net}}}
\newcommand{\mavg}{\langle |m| \rangle}
\newcommand{\btwo}{\left| \beta_2 \right|}
\newcommand{\btwoangle}{\angle{\beta_2}}

\begin{document}

\title{Comparison of Polarized Radiative Transfer Codes used by the EHT Collaboration}

\correspondingauthor{Cora Prather}

\author[0000-0002-0393-7734]{Cora Prather}
\affiliation{CCS-2, Los Alamos National Laboratory, P.O. Box 1663, Los Alamos, NM 87545, USA}

\author[0000-0003-3903-0373]{Jason Dexter}
\affiliation{JILA and Department of Astrophysical and Planetary Sciences, University of Colorado, Boulder, CO 80309, USA}

\author[0000-0002-4661-6332]{Monika Moscibrodzka}
\affiliation{Department of Astrophysics/IMAPP, Radboud University,P.O. Box
             9010, 6500 GL Nijmegen, The Netherlands}

\author[0000-0001-9270-8812]{Hung-Yi Pu}
\affiliation{Department of Physics, National Taiwan Normal University, No. 88, Sec.4, Tingzhou Rd., Taipei 116, Taiwan, R.O.C.}
\affiliation{Center of Astronomy and Gravitation, National Taiwan Normal University, No. 88, Sec. 4, Tingzhou Road, Taipei 116, Taiwan, R.O.C.}
\affiliation{Institute of Astronomy and Astrophysics, Academia Sinica, 11F of Astronomy-Mathematics Building, AS/NTU No. 1, Sec. 4, Roosevelt Rd, Taipei 10617, Taiwan, R.O.C.}

\author[0000-0003-1151-3971]{Thomas Bronzwaer}
\affiliation{Department of Astrophysics/IMAPP, Radboud University,P.O. Box 9010, 6500 GL Nijmegen, The Netherlands}

\author[0000-0002-2685-2434]{Jordy Davelaar}
\affiliation{Department of Astronomy and Columbia Astrophysics Laboratory, Columbia University, 550 W 120th Street, New York, NY 10027, USA}
\affiliation{Center for Computational Astrophysics, Flatiron Institute, 162 Fifth Avenue, New York, NY 10010, USA}
\affiliation{Department of Astrophysics, Institute for Mathematics, Astrophysics and Particle Physics (IMAPP), Radboud University, P.O. Box 9010, 6500 GL Nijmegen, The Netherlands}

\author[0000-0001-9283-1191]{Ziri Younsi}
\affiliation{Mullard Space Science Laboratory, University College London, Holmbury St. Mary, Dorking, Surrey, RH5 6NT, UK}
\affiliation{Institut f\"ur Theoretische Physik, Goethe-Universit\"at Frankfurt, Max-von-Laue-Stra{\ss}e 1, D-60438 Frankfurt am Main, Germany}

\author[0000-0001-7451-8935]{Charles F. Gammie}
\affiliation{Department of Physics, University of Illinois, 1110 West Green Street, Urbana, IL 61801, USA}
\affiliation{Department of Astronomy, University of Illinois at Urbana-Champaign, 1002 West Green Street, Urbana, IL 61801, USA}
\affiliation{NCSA, University of Illinois, 1205 W Clark St, Urbana, IL 61801, USA}

\author[0000-0003-2492-1966]{Roman Gold}
\affiliation{CP3-Origins, University of Southern Denmark, Campusvej 55, DK-5230 Odense M, Denmark}

\author[0000-0001-6952-2147]{George N. Wong}
\affiliation{School of Natural Sciences, Institute for Advanced Study, 1 Einstein Drive, Princeton, NJ 08540, USA}
\affiliation{Princeton Gravity Initiative, Princeton University, Princeton, New Jersey 08544, USA}

\nocollaboration{10}

% ...

\collaboration{500}{The Event Horizon Telescope Collaboration}

\date{\today}

\begin{abstract}
Interpretation of resolved polarized images of black holes by the Event Horizon Telescope (EHT) requires predictions of the polarized emission observable by an Earth-based instrument for a particular model of the black hole accretion system.  Such predictions are generated by general relativistic radiative transfer (GRRT) codes, which integrate the equations of polarized radiative transfer in curved spacetime.  A selection of ray-tracing GRRT codes used within the EHT collaboration is evaluated for accuracy and consistency in producing a selection of test images, demonstrating that the various methods and implementations of radiative transfer calculations are highly consistent.  When imaging an analytic accretion model, we find that all codes produce images similar within a pixel-wise normalized mean squared error (NMSE) of 0.012 in the worst case.  When imaging a snapshot from a cell-based magnetohydrodynamic simulation, we find all test images to be similar within NMSEs of 0.02, 0.04, 0.04, and 0.12 in Stokes $I$, $Q$, $U$, and $V$ respectively.  We additionally find the values of several image metrics relevant to published EHT results to be in agreement to much better precision than measurement uncertainties.
\end{abstract}

\section{Introduction}

In 2019, the Event Horizon Telescope (EHT) collaboration published images of the central black hole in the galaxy M87 (hereafter \bhname), which measured and interpreted the total intensity of radio emission in two bands near 230\,GHz (\citealt{PaperI,PaperII,PaperIII,PaperIV,PaperV,PaperVI}, hereafter \citetalias{PaperI,PaperII,PaperIII,PaperIV,PaperV,PaperVI}).  In 2021, additional results were released measuring the degree and distribution of linear polarization across the image of \bhname, measured via the Stokes parameters $Q$ and $U$ (\citealt{PaperVII,PaperVIII}, hereafter \citetalias{PaperVII,PaperVIII}).  Linear polarization results are also expected of the central-Milky Way black hole \sgra accompanying total-intensity results published in 2022 (\citealt{PaperSI,PaperSII,PaperSIII,PaperSIV,PaperSV,PaperSVI}).

In order to interpret polarized observations, the collaboration generated models of the accreting plasma around \bhname, usually with general relativistic magnetohydrodynamic (GRMHD) simulations.  Simulated images were generated from the models via general relativistic radiative transfer (GRRT) calculations in order to predict the emission visible from earth from the generated plasma state (\citetalias{PaperV}, see also \citealt{Wong2022}).  The total-intensity images produced by various GRRT codes were validated against analytically-defined tests in \citet{Gold2020} and found to be in good agreement.  That paper also compared the output of certain pairs of codes, but not all codes, when imaging GRMHD simulation data.

The interpretation of the linear-polarimetric EHT image, performed in \citealt{PaperVIII} (hereafter \citetalias{PaperVIII}), also used synthetic images.  Polarimetric images are more complicated than total-intensity images as they predict the linear and circular polarization parameters (Stokes $Q$, $U$, $V$) in addition to the total intensity (Stokes $I$). Predicting polarized emission involves solving the coupled polarized radiative transfer equations, which can introduce significant additional computational problems, such as the treatment of rapid Faraday rotation and the need to parallel transport the linear polarization direction through curved spacetime. The additional complexity merits a separate comparison of polarized radiative transfer schemes present in several of the codes compared in \citet{Gold2020}. That comparison is presented in this paper.

This paper provides brief descriptions of the codes compared, specifications of the tests performed, and measurements of code error (where available) or similarity as a group, compared against parameter changes and estimated detector accuracy.

The paper is structured as follows. In Section~\ref{sec:codes} we briefly describe all codes participating in the comparison study. In section~\ref{sec:tests} we define three test problems used to compare the codes. In section~\ref{sec:results} we define metric to evaluate light-curves and image similarities and we present the results of the comparisons. The discussion of the results and limitations of the examined ray-tracing radiative transfer schemes are given in section~\ref{sec:discussion}. We conclude our study in section~\ref{sec:conclusions}.

\section{Participating Codes}
\label{sec:codes}

\subsection{\bhoss}
The \texttt{BHOSS} code \citep{Younsi2012,Younsi2020} numerically integrates, for an arbitrary input spacetime metric tensor, the geodesic equations of motion coupled with the covariant polarized radiative transfer equations.
The solution of the polarized radiative transfer equations is achieved via parallel-propagation of a pair of mutually-orthogonal basis 4-vectors which define the observer's frame.
A fourth-order Runge-Kutta-Fehlberg method with 5th order error estimate and adaptive stepsize control, hereafter RKF4(5), is typically used.
In regions of higher Faraday depth, a RKF8(9) method is used, and when the transfer equations are particularly stiff a variable-order implicit RKF integrator is employed.

\subsection{\tt ipole}

The \ipole code\footnote{Current version is available at \url{https://github.com/moscibrodzka/ipole}}  (\citealt{Moscibrodzka2018}, \citealt{Noble2007}) is a publicly available ray-tracing scheme for covariant polarized GRRT. \ipole splits the radiative transfer problem into two steps. In the fluid frame it evolves the Stokes parameters taking into account synchrotron emission, absorption and Faraday effects and using an analytic solution to the polarized transfer equations with constant coefficients. Currently two analytic solvers are implemented in the code (\citealt{LandiDegl'Innocenti1985} and the one presented in the Appendix A of \citealt{Moscibrodzka2018}). Analytic solvers make \ipole 
solutions numerically stable even for plasma with large optical or Faraday depths. In the coordinate frame, the parallel transport of the Stokes parameters is accomplished by transport of coherency matrix rather than Stokes parameters themselves. Hence \ipole radiative transfer is coordinate and metric independent. \ipole has been tested against another polarized ray-tracing code \grtrans (see method paper \citealt{Moscibrodzka2018} and the next subsection) and against a polarized Monte Carlo radiative transfer scheme (see \citealt{Moscibrodzka2020} and Appendix~\ref{app:sed} of this work). \ipole was used in \citetalias{PaperVIII} for calculating polarized images of models with accelerated electrons. 

\subsection{\ipoleIL}

\ipoleIL\footnote{\url{https://github.com/AFD-Illinois/ipole}}, usually also called \ipole but suffixed in this comparison for clarity, is a fork of the original \ipole code described above, with features designed for treating libraries of GRMHD snapshot files, particularly from \iharm as a part of the {\tt{}PATOKA} pipeline \citep{Wong2022}.  It maintains the same transport scheme implemented in \ipole, but adds robustness features such as reorthogonalization of tetrad basis vectors and additional limiting cases for the analytic solutions and fits. It also adds compatibility with a number of different GRMHD codes and supports calculating emission from different electron energy distribution functions.

\subsection{\grtrans}

The \texttt{\grtrans} code\footnote{\url{https://github.com/jadexter/grtrans}}  \citep{Dexter2009, Dexter2016} solves the polarized radiative transfer equations along null geodesics in a Kerr spacetime. The radiative transfer equations are integrated either numerically \citep{Hindmarsh2019} or with quadrature methods \citep{LandiDegl'Innocenti1985,Rees1989}. The quadrature methods are the most accurate and efficient for calculations of polarized radiative transfer in Faraday thick problems and are used for the test problems discussed here. 

\subsection{\odyssey}
The \odyssey code \citep{Pu2016} is a public GPU-based code\footnote{\url{https://github.com/hungyipu/Odyssey}} which solves the unpolarized radiative transfer equation along null geodesics from the observer to the source (observer-to-source) in Kerr spacetime. In \citet{Pu2018}, to implement the polarization computations and fit the need for solving the Stokes parameters along the null geodesic from the source to the observer (source-to-observer), a two stage scheme is proposed: (i) during the observer-to-source stage,  unpolarized radiative transfer is computed backward in time, (ii) inverse the time direction and trace the same geodesic during source-to-observer stage, and simultaneously solve the four Stokes parameters. As a result, there are four additional ODEs (related to the Stokes parameters) to be solved in the second stage compared to that in the first stage. By controlling the time direction directly in the code, there is no need to save the photon path during the observer-to-source stage for the use of source-to-observer stage. However, the caveat is that the cost for solving additional four Stokes parameters during the source-to-observer stage can be computationally costly, and the Runge-Kutta scheme may fail when complicated Faraday coefficients are introduced in a given problem. 

In this work, to improve its speed and the stability, we improve the polarization scheme of \odyssey
with the following: (i) a two stage scheme is still adopted, without solving the four ODEs for Stokes parameters during the second (source-to-observer) stage. (ii) Instead, during the second stage, the Stokes parameters are solved along the geodesics with an implicit method \citep{Bronzwaer2020, Pihajoki2018}. In this new scheme, the accuracy of the polarization computation is automatically controlled by the accuracy of the geodesic computation. The modifications significantly improve the computational speed. For example, it takes about a second (including the time for reading GRMHD simulation data) for \odyssey to finish the computation for the GRMHD snapshot test problem (\S 3.3).

\subsection{\raptor}

The \raptor code \citep{Bronzwaer2018, Bronzwaer2020} \footnote{\url{https://github.com/jordydavelaar/raptor}} is a public code that numerically integrates the equations of motion of light rays in arbitrary spacetimes and then performs polarized radiative transfer calculations along the rays. The code uses an adaptive Runge-Kutta-Fehlberg scheme to integrate the geodesic equation where the Christoffel symbols can either be provided analytically or are numerically computed on the fly by using a fourth order centered finite difference method. To integrate the polarized radiative transfer equation \raptor uses a hybrid ImEx integration scheme that switches to an implicit integrator in case of stiffness, in order to solve the equation with optimal speed and accuracy for all possible values of the local optical/Faraday thickness of the plasma. The code uses an adaptive camera grid to optimize run time by adding resolution where needed \citep{davelaar2022}, and can produce virtual reality visualizations \citep{davelaar2018}. The code is fully interfaced with the non-uniform grid (adaptive mesh refinement) data format of the {\tt BHAC} code \citep{davelaar2019}. Radiative transfer coefficients are provided for the thermal electron distribution, but also the $\kappa$ and power-law distributions.

\section{Test Problems}\label{sec:tests}

Three test problems were used to evaluate the codes.  The problems were chosen to reflect tests already present in the literature, highlighting specifically the aspects of code performance related to polarized transport.  A previous comparison \citep{Gold2020} evaluated many of the same codes for similarity and accuracy in producing total-intensity images.  An additional goal was to verify data product similarity when imaging the output of GRMHD simulations, a test evaluated only for certain pairs of codes considered in \citet{Gold2020}.  The tests are described here from least to most complex, with each testing a larger subset of code features.

\subsection{Comparison to Analytic Result}
\label{sec:prob_analytic}

The first test problem is a straightforward integration of the non-relativistic polarized transfer equation using constant coefficients, chosen for the availability of an analytic solution from \citet{LandiDegl'Innocenti1985}, allowing direct evaluation of code accuracy, in addition to code similarity.  The test here is taken directly from \citet{Dexter2016}, with coefficients as listed in \citet{Moscibrodzka2018}.

In the Stokes basis $I$, $Q$, $U$, $V$, the non-relativistic polarized radiative transfer equation is
\begin{equation}
\frac{d}{d s}
\begin{pmatrix} I \\ Q \\ U \\ V \end{pmatrix}
 = \begin{pmatrix} j_{I} \\ j_{Q} \\ j_{U} \\ j_{V} \end{pmatrix}
- \begin{pmatrix} 
\alpha_{I} & \alpha_{Q} & \alpha_{U} & \alpha_{V} \\
\alpha_{Q} & \alpha_{I} & \rho_{V} & -\rho_{U} \\
\alpha_{U} & -\rho_{V} & \alpha_{I} & \rho_{Q} \\
\alpha_{V} & \rho_{U} & -\rho_{Q} & \alpha_{I} 
\end{pmatrix}
\begin{pmatrix} I \\ Q \\ U \\ V \end{pmatrix} .
\end{equation}

In the test, this equation is integrated twice with different subsets of coefficients nonzero.  This minimizes the complexity of the analytic comparison functions, isolating any bugs in treating emission and absorption from those in Faraday rotation and conversion.  The coefficients for each integration are given in Table \ref{tab:analytic_values}.

\begin{deluxetable*}{l|cccc|cccc|ccc}
    \tablehead{ \colhead{} & \colhead{$j_I$} & \colhead{$j_Q$} & \colhead{$j_U$} & \colhead{$j_V$} &
    \colhead{$\alpha_I$} & \colhead{$\alpha_Q$} & \colhead{$\alpha_U$} & \colhead{$\alpha_V$} &
    \colhead{$\rho_Q$} & \colhead{$\rho_U$} & \colhead{$\rho_V$}}
    \startdata
    Emission/Absorption & 2.0 & 1.0 & 0.0 & 0.0 & 1.0 & 1.2 & 0.0 & 0.0 & 0.0 & 0.0 & 0.0 \\
    Rotation            & 0.0 & 0.1 & 0.1 & 0.1 & 0.0 & 0.0 & 0.0 & 0.0 & 10.0 & 0.0 & -4.0
    \enddata
    \caption{Constant coefficients for the analytic comparison test integrations -- these mirror values in \citet{Moscibrodzka2018}. See Section \ref{sec:prob_analytic}.}
    \label{tab:analytic_values}
\end{deluxetable*}

\subsection{Thin-Disk Model}
\label{sec:prob_thindisk}

The second test problem consists of imaging emission from a thin opaque disk aligned to the midplane of a near-maximally spinning black hole, as described in \citet{Novikov1973}.  This involves solving the geodesic equation in two contexts: first in tracing lines of sight from the camera through the Kerr metric and second in parallel-transporting the direction of linearly polarized emission from the disk back to the camera.  This test closely mirrors a figure from \citet{Schnittman2009}, which is reproduced as a test in \citet{Dexter2016}.

The thin-disk test does not include any diffuse emission: that is, all transport coefficients $j_S, \alpha_S, \rho_S = 0$ uniformly for all Stokes parameters S.  Instead, the initial Stokes parameters are set as a boundary condition at the first midplane crossing of each geodesic when traced backward from the camera.  As in \cite{Schnittman2009}, the total flux $F$ is taken from \cite{Page1974}, and the intensity at the desired frequency $I_{\nu}$ is obtained by calculating an effective temperature and assuming a black-body distribution diluted by a hardening factor $n = 1.8$:
\begin{align}
    T_{\mathrm{eff}} &\equiv \left( \frac{F}{\sigma} \right)^{1/4}, \\
    I_{\nu} &= \frac{1}{n^4} B_{\nu} (n \cdot T_{\mathrm{eff}}),
\end{align}
where $F$ is the total emitted power per area of the thin disk, and $B_{\nu}(T)$ is the black-body function of the temperature $T$ and the emitted frequency $\nu$ in the fluid frame.

The emitted intensity and horizontal polarization fraction in the outgoing direction are determined by assuming scattering from a semi-infinite atmosphere, as in \cite{Chandrasekhar1960}, Table 24, with the direction of linear polarization pointing along the plane of the disk.  Emission is enabled only between $r_{\mathrm{ISCO}}$ and $R_{\mathrm{out}} = 100 r_g$, where $r_g$ is the system gravitational radius $G M_\mathrm{BH} / c^2$ with $G$ the gravitational constant and $c$ the speed of light.  The fluid orbital angular velocity $\nicefrac{u^\phi}{u^t}$ is assumed to be Keplerian:
\begin{align}
    \frac{u^\phi}{u^t} = \frac{1}{r^{3/2} + \bhspin}.
\end{align}
where $\bhspin$ is the dimensionless form of the BH angular momentum $J$, $\bhspin \equiv J c / {G M^2}$ with $-1 \leq \bhspin \leq 1$.
As the test will need to be implemented in many different codes, we simplify the original problem from \citet{Schnittman2009} by observing at only a single frequency rather than summing over a range.  The full set of parameters used for this image is:
\begin{subequations}
\begin{align}
    \bhspin &= 0.99 \\
    M_{\mathrm{BH}} &= 10 \; M_{\odot} = 1.477 \cdot 10^6 \mathrm{cm} \frac{c^2}{G} \\
    D_{\mathrm{source}} &= 0.05 \; \mathrm{pc}\\
    \dot{M} &= 0.1 \; \dot{M}_{\mathrm{Edd}} \approx 2.218 \cdot 10^{-9} M_{\odot} / \mathrm{yr} \\
    h \nu &= 1 \mathrm{keV}
\end{align}
\end{subequations}
where these parameters define observation at a single frequency $\nu$ of a BH of mass $M_{\mathrm{BH}}$ at distance $D_{\mathrm{source}}$, characterized by a Kerr spacetime with BH spin parameter $\bhspin$, and accreting at rate $\dot{M}$.  Following Dexter and Schnittman, $\dot{M}_{\mathrm{Edd}} \equiv \nicefrac{L_{\mathrm{Edd}}}{c^2}$.

\begin{figure}
    \centering
    \includegraphics[width=0.5\textwidth]{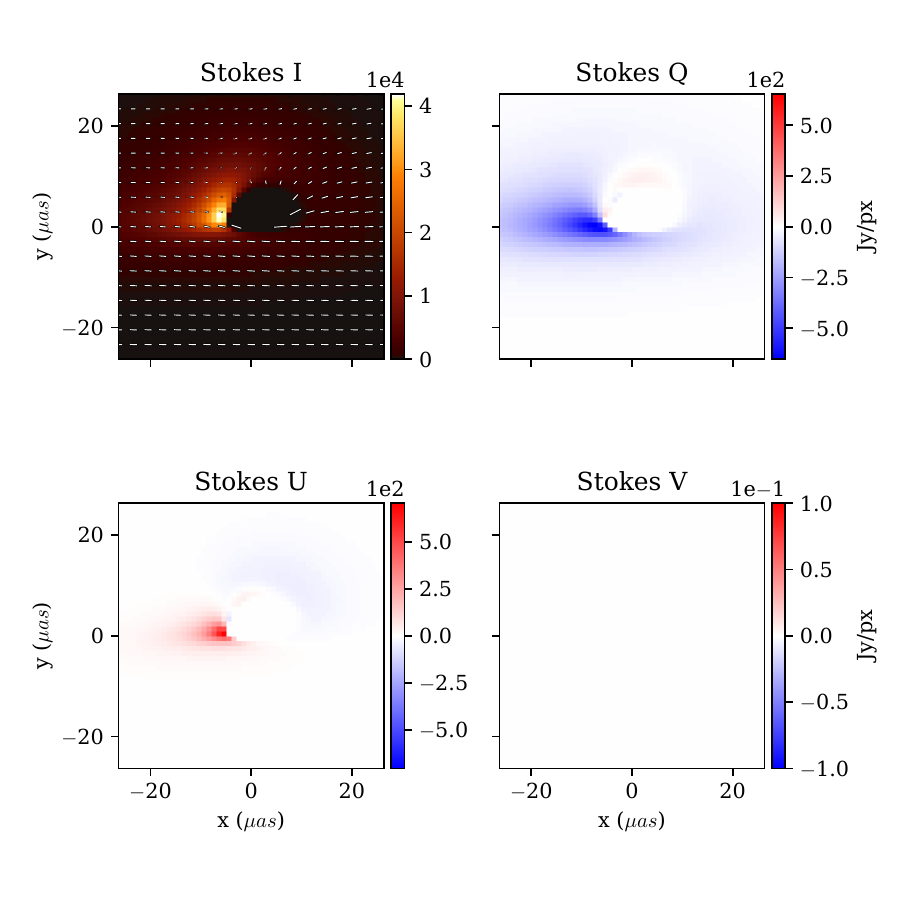}
    \caption{Example output of the thin-disk problem: the upper left panel shows total flux with overplotted polarization direction vectors scaled by the linear polarization fraction, and the other panels show fluxes of Stokes $Q$, $U$, and $V$ at each pixel (the Stokes $Q$, $U$, and $V$ images). In this test Stokes $V$ remains identically zero as expected.}
    \label{fig:prob_thindisk}
\end{figure}

Example output from this test run with \ipoleIL is shown in Figure \ref{fig:prob_thindisk}.  As no circularly polarized emission or Faraday conversion occurs in the problem, the Stokes V flux remains exactly zero.

\subsubsection{Camera}
\label{sec:prob_camera}
In this and the following test, the camera tetrad is constructed such that a geodesic at the center of the camera's field of view (FOV) would have zero angular momentum $k_{\phi}$.  The polar angle $\theta_{\mathrm{cam}}$ is defined relative to the BH angular momentum vector.  This is identical to the camera definition from \citet{Gold2020}.

The camera is placed at radius $10^4\,r_g$ in both tests, to reduce the discrepancy between pinhole and planar cameras.  Note that $R_{\mathrm{cam}}$ is in this case and in practice much smaller than $D_\mathrm{source}$ -- so long as $R_{\mathrm{cam}}$ is large enough to eliminate camera effects, the image intensity is invariant with distance.

The FOV of each test is given in two forms: DX, the in-plane distance from one edge of the imaged material to the other in gravitational radii $r_g$, and FOV, the angular size from Earth in micro-arcseconds ($\mu\mathrm{as}$).

Pixels on each image correspond to evenly spaced geodesics, starting from each pixel center.  That is, an image with a FOV of 80 $\mu\mathrm{as}$ to a side and NX of 80 pixels would be calculated using geodesics originating at -39.5 $\mu\mathrm{as}$ to 39.5 $\mu\mathrm{as}$ away from the FOV center, spaced 1 $\mu\mathrm{as}$ apart in each cardinal direction.  All images in these tests are square, with equal FOV from West to East and North to South.

For the thin-disk test, the camera parameters are:
\begin{subequations}
\begin{align}
    \theta_{\mathrm{cam}} &= 75^{\circ} \\
    \mathrm{NX} &= 80 \, \mathrm{px} \\
    \mathrm{DX} &= 40 \, r_g \\
    \mathrm{FOV} &= 78.99 \, \mu\mathrm{as}
\end{align}
\end{subequations}

\subsection{GRMHD Snapshot}
\label{sec:prob_grmhd}

The last test consists of imaging the relativistic thermal synchrotron emission at 230GHz from one snapshot from a GRMHD simulation. This test exercises all aspects of the code as well as code-specific choices, such as interpolation of fluid state recorded at discrete locations and calculation of the transport coefficients via fitting functions.  The standard snapshot file used for this test is taken from a SANE simulation with spin $\bhspin = 0.9375$, performed using \iharm \citep{Prather2021} with a resolution of 288x128x128 cells in $r$, $\theta$, and $\phi$ respectively.  Except for the coordinate system, this simulation exactly reflects the simulations performed with \iharm as a part of the library used in \citetalias{PaperV} and \citetalias{PaperVIII}, as described in \citet{Wong2022}.  Not all GRRT codes can read all GRMHD output, as coordinate systems and fluid state descriptions can differ from code to code -- thus, not every polarized radiative transfer code used in the EHTC can be directly compared with this test.  The \iharm format is chosen as it is readable by a majority of codes used in the EHTC, and in particular those codes relevant to studies in \citetalias{PaperVIII}.

The snapshot is taken at $4500\, r_g/c$ after simulation start, well into the run's quiescent period.  The file is available upon request for testing future codes.  Since it involves creating an image from just one snapshot of the simulation, the test makes the assumption of ``fast-light,'' i.e.,~that the fluid is static as light propagates from emission to observer.

The parameters of this test are chosen to reflect values for \bhname, specifically those used in creating the libraries of simulated images used in \citetalias{PaperV}, \citetalias{PaperVI}, and \citetalias{PaperVIII}, hereafter the ``EHT image libraries.''  These are:
\begin{subequations}
\begin{align}
    M_{\mathrm{BH}} &= 6.2 \times 10^9 \; M_{\odot} \\
    D &= 16.9 \, \mathrm{Mpc} \\
    \nu &= 230 \, \mathrm{GHz}
\end{align}
\end{subequations}
The camera is defined as in section \ref{sec:prob_camera}, with parameters chosen to reflect the angle of the \bhname jet and the FOV observed by the EHT:
\begin{subequations}
\begin{align}
    \theta_{\mathrm{cam}} &= 163^{\circ} \\
    \phi_{\mathrm{cam}} &= 0^{\circ} \\
    \mathrm{FOV} &= 160 \, \mu\mathrm{as} \\
    \mathrm{DX} &= 44.17 \, r_g \\
    \mathrm{NX} &= 160 \, \mathrm{px}
\end{align}
\end{subequations}

In addition to the system parameters above, imaging a GRMHD simulation necessarily involves setting another scale factor which determines the density of accreting material and the strength of magnetic fields.  It is expressed here as a mass unit, $\mathcal{M}$, which gives units to the unscaled density values from a simulation, $\rho_{\mathrm{code}}$:
\begin{align}
\rho_{\mathrm{CGS}} = \frac{\mathcal{M}}{r_g^3} \rho_{\mathrm{code}}
\end{align}
$\mathcal{M}$ is not known a priori, but it is highly correlated with the total image brightness.  Thus, it is scaled so as to match the total image flux density to the observed compact flux density, usually by employing an iterative solver.

In the EHT image libraries, $\mathcal{M}$ was fit such that images taken over the course of a full simulation would produce an average of $0.5$ Jy of compact flux density (see \citealt{Wong2022} for details).  For this test, the sample image is fit alone using \ipoleIL such that it produces $0.50$ Jy when imaged with totally unpolarized transport, or about $0.47$ Jy of Stokes $I$ flux density when imaged using polarized transport.  The value of $\mathcal{M}$, and the corresponding accretion rate $\dot{M}$, used for the test are listed below:
\begin{subequations}
\begin{align}
    \mathcal{M} &= 1.672 \cdot 10^{26} \;  \mathrm{g} \\
    \dot{M} &=  8.644 \cdot 10^{-5} \;  M_{\odot} / \mathrm{yr} =  6.285 \cdot 10^{-7} \; \dot{M}_{\mathrm{Edd}}
\end{align}
\end{subequations}
where $\dot{M}_{\mathrm{Edd}}$ is defined as earlier in this work.

Accretion flows around \bhname are strongly suspected to be two-temperature, with little thermal coupling between the ions and electrons \citep{Mahadevan1997,Ryan2017,Sadowski2017}.  Since GRMHD simulations evolve only a single fluid with a single temperature, when simulating images the internal energy must be split between the ions and electrons based on a model.  While this process is not well constrained, it is generally documented which electron distribution is being assumed, and the electron energy distribution model is standardized between codes when similar performance is expected.

Thus, for simplicity, in this test we set the electron temperature to a fixed ratio of $1/3$ of the ion temperature, derived from the single-fluid GRMHD parameters by holding the total internal energy constant (see \citetalias{PaperV}):
\begin{align}
    T_e = \frac{2 m_p u}{15 k \rho}
\end{align}
where $u$ and $\rho$ are the local fluid internal energy and rest-mass density per unit volume, respectively, and $m_p$ and $k$ are the proton mass and Boltzmann constant.  Note that this is equivalent to the so-called ``$\rhigh$'' model of \citetalias{PaperV} with $\rlow = \rhigh = 3$.  In splitting the total internal energy rather than setting the fluid and ion temperatures equal, it differs slightly from the original statement of the $\rhigh$ model in \cite{Moscibrodzka2016}.

The emission, absorption, and rotation coefficients are calculated based on the electron temperature (or more broadly, the electron energy distribution) using fitting functions approximating the full synchrotron emission calculations, which are expensive to compute.  Codes in this comparison used a few different sets of fitting functions; further discussion is found in Section \ref{sec:caveats} and Appendix \ref{app:coefficients}.

Finally, as in \citetalias{PaperV} and commonly in the literature, emission is tracked only from regions of the simulation with $\sigma \equiv B^2/\rho < 1$.  Regions with higher sigma (largely the polar ``jet'' regions) can over-produce emission if included, due to hot material in the jet inserted by numerical floors to preserve stability of GRMHD algorithms.

\begin{figure}
    \centering
    \includegraphics[width=0.45\textwidth]{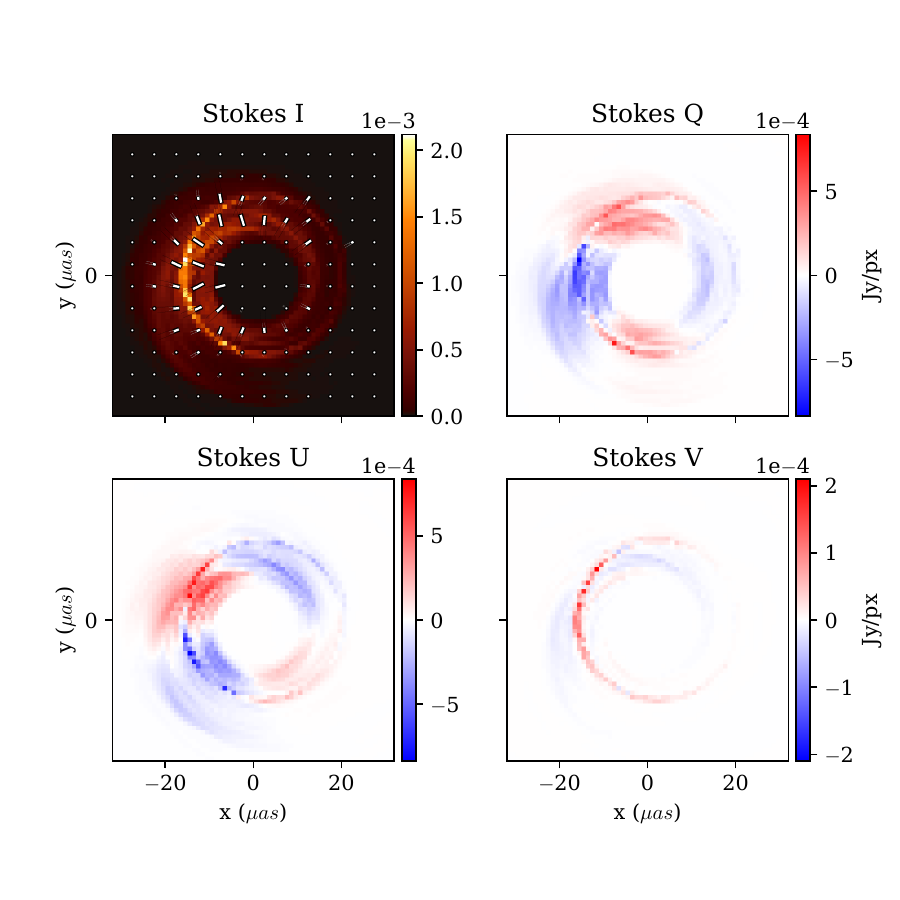}
    \caption{Example output from \ipoleIL running the GRMHD snapshot test.  The image was produced using the parameters listed in Section \ref{sec:prob_grmhd}, with the accretion rate parameter $\mathcal{M}$ fit so as to produce about $0.5$ Jy of total flux density at 230 GHz to match EHT observations.}
    \label{fig:stokes_sample}
\end{figure}

Example output for the GRMHD snapshot test from \ipoleIL is provided in Figure \ref{fig:stokes_sample}.

\section{Results}\label{sec:results}

\subsection{Analytic Comparison Results}
\label{sec:results_analytic}

\begin{figure*}
    \centering
    \includegraphics[width=0.9\textwidth]{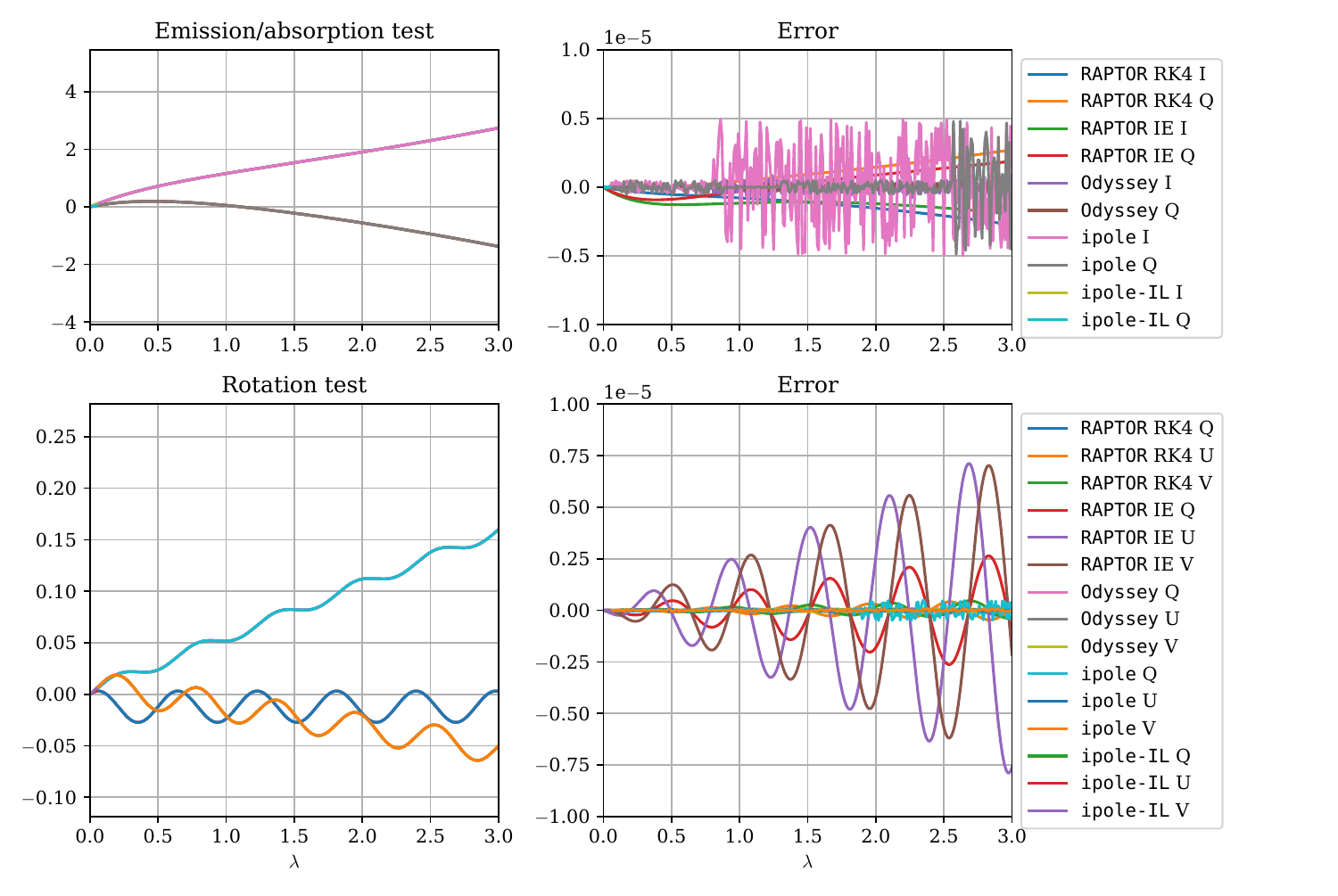}
    \caption{Comparison of integrator results for the analytic tests, plotting the relevant Stokes parameters against the dimensionless affine parameter $\lambda$, equivalent to length in this non-relativistic test.  Output from all codes overlaps to within line widths in the left panes.}
    \label{fig:analytic_result}
\end{figure*}

Results for the analytic integration tests from \raptor, \odyssey, \ipole, and \ipoleIL are shown in Figure \ref{fig:analytic_result}. Raw output is plotted in the left panes, and differences from the analytic result are plotted on the right.  This test verifies that the default accuracy parameters of each code allow them to match an analytic solution to within acceptable errors.  Note that this is not a good measure of relative code accuracy or convergence---for convergence tests, see the accompanying code papers cited in Section \ref{sec:codes}.

Note that the results from two integrators are shown for \raptor as the ``RK4'' and ``IE'' variants. In normal integration, \raptor uses the ``RK4'' integrator, reserving the ``IE'' integrator for the few zones where Faraday rotation is too strong to take steps of an appropriate size with an explicit scheme \citep{Bronzwaer2020}.

Also, the \ipole scheme (also used in \ipoleIL) is semi-analytic: it uses the analytic solution for constant coefficients whenever it evolves the non-relativistic polarized radiative transfer equations.  Thus, \ipole and \ipoleIL will perform this test \emph{less} accurately when taking more steps: a single step of any size would be exactly accurate, but multiple steps accrue round-off error.

\subsection{Metrics}
\label{sec:metrics}

In the following two imaging tests, no exact result is available by which to evaluate code accuracy directly; rather, we evaluate consistency between all codes, both in overall image structure and in several metrics used to compare models with EHT results.  In particular, we will use the definitions from \citetalias{PaperVIII}, computed over simulated images and used to compare models to the observed EHT result.  These summary statistics include the total flux density $F$, image-integrated or ``zero-baseline'' linear and circular polarization fractions $\mnet$ and $\vnet$, and the average linear polarization fraction over the resolved image $\mavg$.  For an image represented as a vector of emitted flux per pixel in each Stokes parameter $I_j, Q_j, U_j, V_j$ over each pixel $j$, these values are defined as:

\begin{align}
    \mathrm{F} &= \sum_j I_j \\
    \mnet &= \frac{\sqrt{ \left(\sum_j Q_j \right)^2 + \left(\sum_j U_j \right)^2}}{\sum_j I_j} \\
    \mavg &= \frac{\sum_j \sqrt{Q_j^2 + U_j^2}}{\sum_j I_j} \\
    \vnet &= \frac{\sum_j V_j}{\sum_j I_j}
\end{align}

Additionally, \citetalias{PaperVIII} used a complex coefficient reflecting the degree and angle of azimuthally symmetric linear polarization, $\beta_2$ (see \citealt{Palumbo2020}).  It is calculated by first centering the image, e.g.,~with {\tt rex} \citepalias{PaperIV}, and then taking the inner product of the complex linear polarization $P = Q + iU$ with a rotationally symmetric function:
\begin{align}
    &\beta_{2} = \frac{1}{I_{\mathrm{ann}}} \int_{\rho_{\mathrm{min}}}^{\rho_{\mathrm{max}}} \int_0^{2 \pi} P(\rho, \phi) e^{-i 2 \phi} \; \rho \, d\phi d\rho \\
    &I_{\mathrm{ann}} = \int_{\rho_{\mathrm{min}}}^{\rho_{\mathrm{max}}} \int_0^{2 \pi} I(\rho, \phi) \; \rho \, d\phi d\rho,
\end{align}
where $\rho$/$\phi$ are polar coordinates in the image plane, measured from/about the image center.  This metric is expected to be useful only for images with a relatively low observer angle, as they will be more symmetric; thus it is computed and compared only for the GRMHD snapshot test, which uses the low observer angle $i = 17^\circ$ expected for \bhname and used for libraries of simulated images of that object.

The quantities $\mavg$ and $\beta_2$ are sensitive to image resolution.  In order to mirror EHT measurements and reflect how simulated images were used in \citetalias{PaperVIII}, a circular Gaussian blur with a FWHM of 20$\mu\mathrm{as}$ was applied to all images before computing either resolution-dependent quantity.

In addition to the quantities used for direct comparison in \citetalias{PaperVIII}, we measure a point-source linear polarization direction or electric vector position angle (EVPA) East of due North on the sky, and thus in our Stokes convention defined as:
\begin{align}
    \mathrm{EVPA}  &= \frac{1}{2} \mathrm{arg}{\left( \frac{\sum_j U_j}{\sum_j Q_j} \right)}.
\end{align}
As this metric is potentially volatile for images with low net linear polarization, and a corresponding measurement has not been made to which we might compare, we follow \citetalias{PaperVIII} in omitting this as a comparison metric -- rather, we use it only in image summaries.

When evaluating image similarity, we use the normalized mean squared error (NMSE):

\begin{align}
    \mathrm{NMSE}(A, B) &= \frac{\sum_j|A_j-B_j|^2}{\sum_j|A_j|^2},
\end{align}
where $A_j$ and $B_j$ are the intensities of a particular Stokes parameter in two images at pixel $j$. Regardless of exact nomenclature, all mean squared error values listed in this work and in \citet{Gold2020} use this normalization and are thus comparable.

Note that this definition of the NMSE is not symmetric under the ordering of $A$ and $B$, and in particular as images get dimmer, $\mathrm{NMSE}(I, 0) = 1$ whereas $\mathrm{NMSE}(0, I) = \infty$. As it is normalized against the sum of squared pixel intensities, the NMSE becomes more volatile when evaluating dimmer images.

The NMSE was one of two metrics used to gauge similarity in \citet{Gold2020}.  We omit the other, the structural dissimilarity (DSSIM), since for the case of very similar images, values of the DSSIM are highly correlated with the NMSE (see Appendix \ref{app:metric_compare}).

\subsection{Thin Disk Test Results}
\label{sec:results_thindisk}

\begin{figure}[H]
    \centering
    \includegraphics[width=0.5\textwidth]{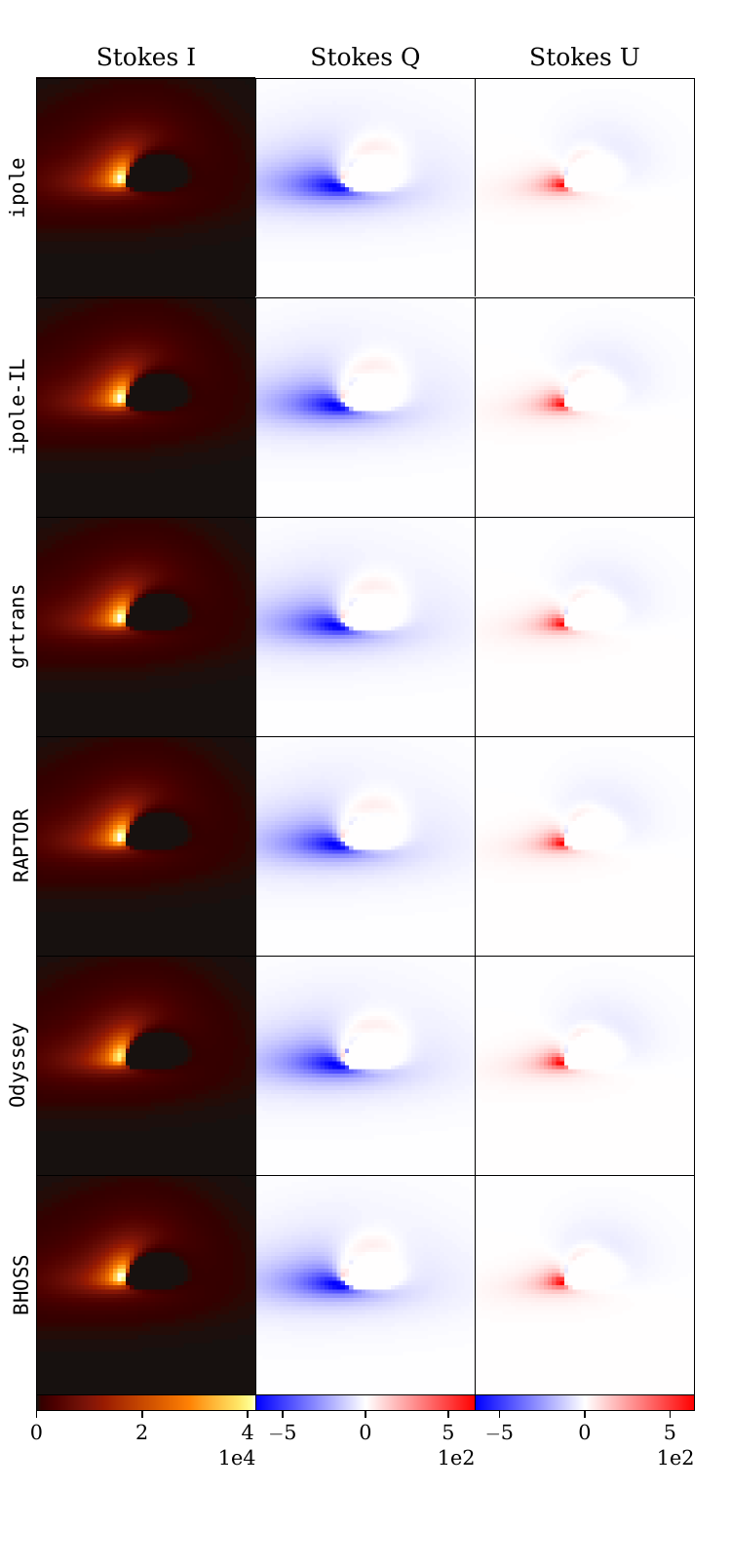}
    \caption{Full set of images produced for the thin disk problem, in which codes produced an image from an analytic prescription for an opaque thin disk.  The Stokes parameters $I$, $Q$, $U$ are plotted separately, with Stokes $V$ omitted as it is uniformly zero.}
    \label{fig:thin_disk_images}
\end{figure}

Each code's output for the thin disk test is plotted in Figure \ref{fig:thin_disk_images}.  The images are visually indistinguishable except in a few particular pixels, and this similarity is borne out in the comparison metrics.  Stokes V is omitted from plots and comparisons for this test, as all codes produce exactly zero Stokes $V$ across the entire image.

\begin{figure*}
    \centering
    \includegraphics[width=\textwidth]{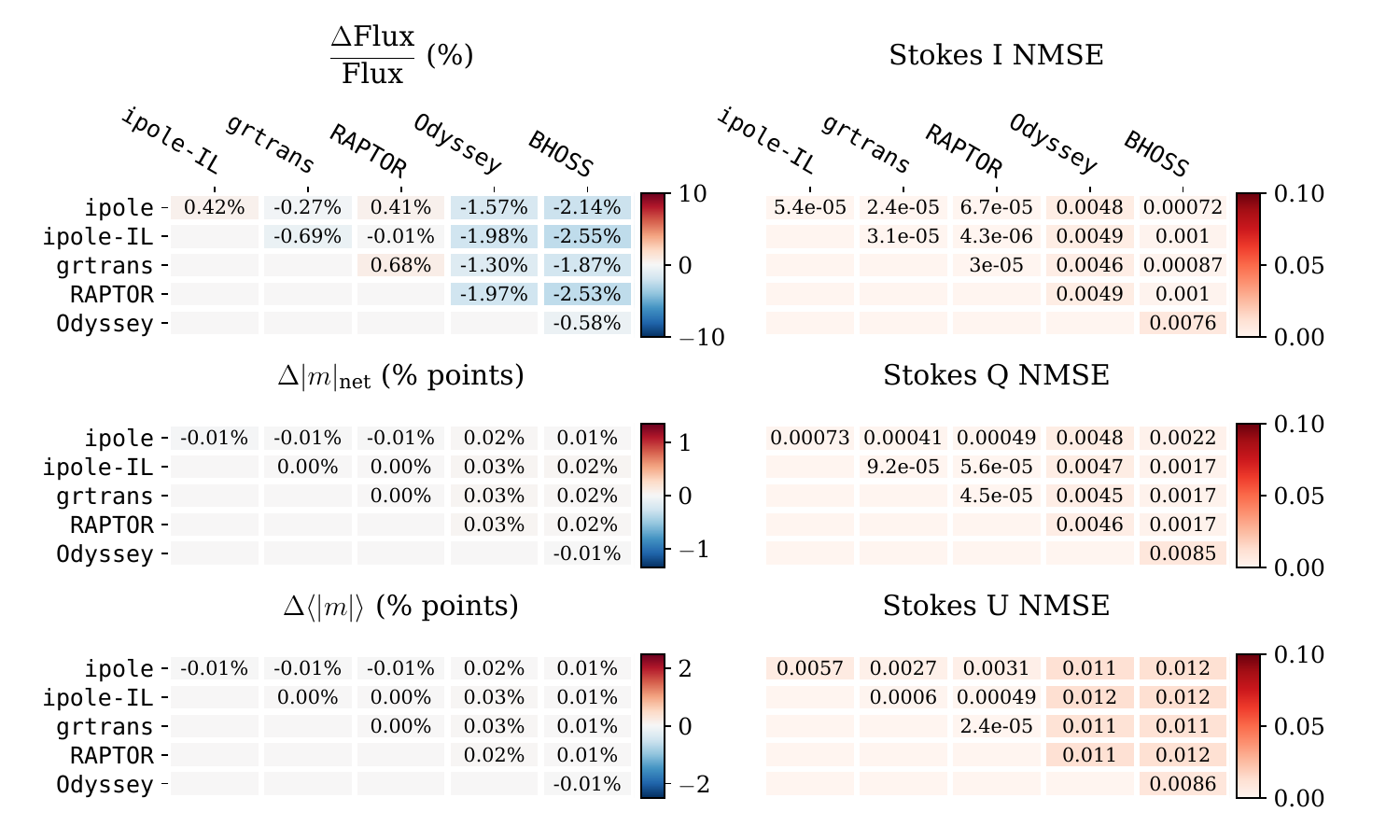}
    \caption{(left column) Tables comparing the absolute differences between the images produced by each pair of codes running the thin disk test.  The values themselves are provided in Table \ref{tab:thin_disk_int}.  Since the differences are symmetric, the table takes only the upper triangular portion of the comparison.
    (right column) Tables comparing the normalized mean squared error (NMSE) between each pair of images, as defined in Section \ref{sec:metrics}.
    To aid in comprehension, table cells are colored by value.  Circular polarization is omitted, see test description in Section \ref{sec:prob_thindisk}.}
    \label{fig:thin_disk_tables}
\end{figure*}

Recall that this test involves accurate evaluation of the geodesic equation and accurate parallel transport of the linear polarization vector from emission to camera. As the codes' similarity in tracing geodesics was evaluated extensively in \citet{Gold2020}, we focus here on demonstrating the latter through comparison of the resulting Stokes $Q$, $U$ images and the relevant image-integrated metrics.  Table \ref{tab:thin_disk_int} lists total fluxes and net polarization parameters for each image in the test.  Figure \ref{fig:thin_disk_tables} presents comparisons of each metric between each pair of images.

\begin{deluxetable}{l|cccc}
    \tablehead{ \colhead{Code } & \colhead{Flux [Jy]} & \colhead{$\mnet$ [\%]} & \colhead{$\mavg$ [\%]} & \colhead{EVPA [$^\circ$]} }
    \startdata
\texttt{ipole} & 6.841e+06 & 2.3341 & 2.378 & 88.123 \\
\texttt{ipole-IL} & 6.8699e+06 & 2.3224 & 2.3707 & 87.974 \\
\texttt{grtrans} & 6.8227e+06 & 2.3246 & 2.3709 & 88.01 \\
\texttt{RAPTOR} & 6.8689e+06 & 2.3265 & 2.3726 & 88.025 \\
\texttt{Odyssey} & 6.7338e+06 & 2.3527 & 2.3971 & 88.146 \\
\texttt{BHOSS} & 6.6949e+06 & 2.3439 & 2.3845 & 88.131
    \enddata
    \caption{Image-integrated values for the thin disk test.  $\vnet$ is omitted as it is uniformly zero, and $\beta_2$ magnitude and angle are omitted as the image is not symmetric and thus the magnitude is very small.}
    \label{tab:thin_disk_int}
\end{deluxetable}

\subsection{GRMHD Snapshot Test Results}
\label{sec:results_grmhd}

\begin{figure*}
    \centering
    \includegraphics[width=\textwidth]{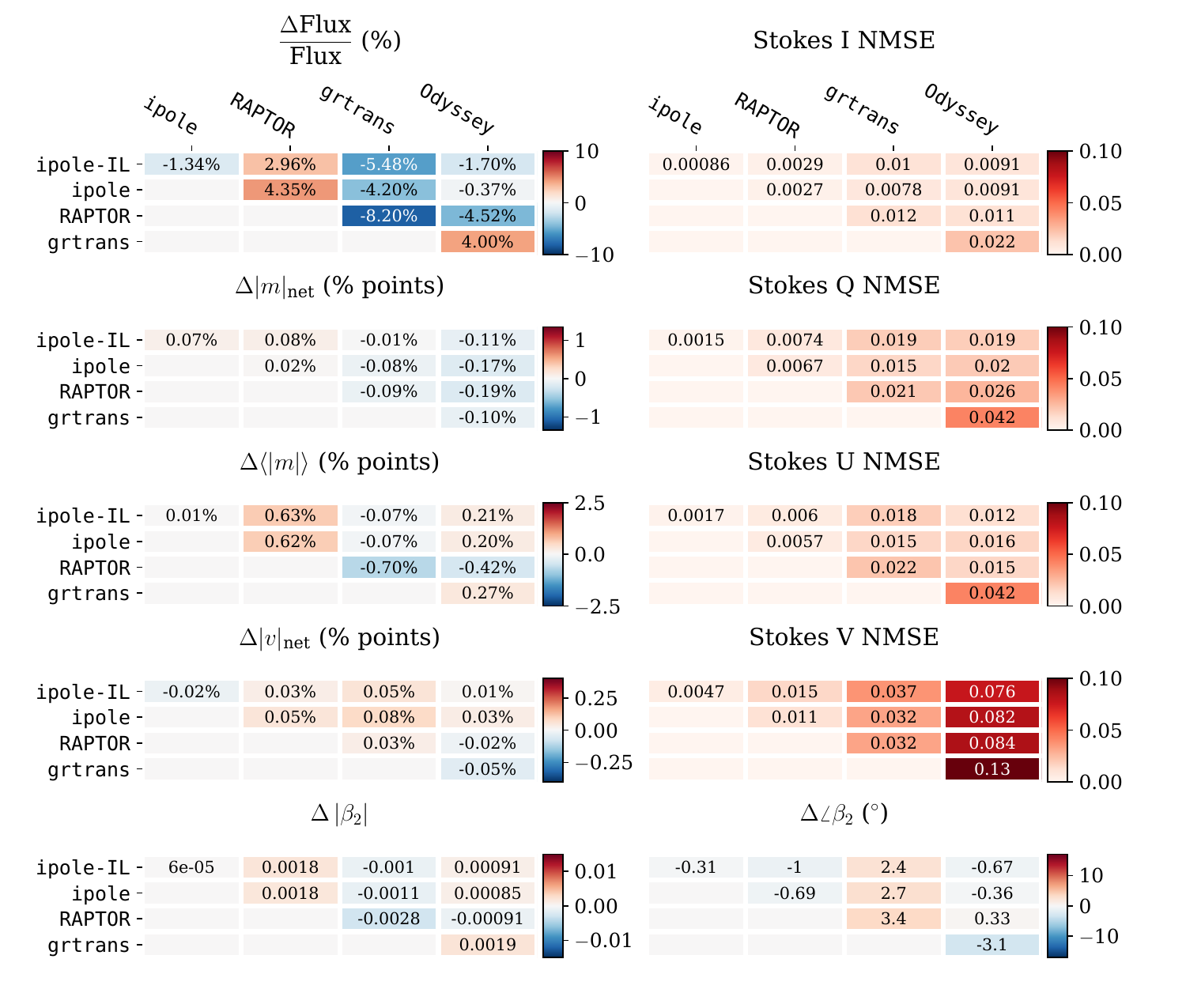}
    \caption{Tables comparing each pair of images in the GRMHD snapshot test.  (left column, bottom row) Tables comparing all six image metrics used in \citetalias{PaperVIII}. Note that except for total flux density, all comparisons are absolute: that is, images with $\vnet$ of 0.4\% and 0.8\% would be listed with a difference of 0.4\% points.  (right column) Tables listing the NMSE between each pair of images.  See Section \ref{sec:metrics} for the definitions of all comparison metrics and Figure \ref{fig:thin_disk_tables} for a description of table layout.}
    \label{fig:grmhd_hi_tables}
\end{figure*}

Results for the GRMHD snapshot test for \ipole, \ipoleIL, \grtrans, \odyssey, and \raptor are listed in Table \ref{tab:grmhd_hi_int} and presented in Figure \ref{fig:grmhd_hi_tables}.  Results are formatted similarly to the results of the thin disk test, with the addition of a Stokes V component and circular polarization fraction in images and tables and the addition of the rotationally symmetric linear polarization coefficient $\beta_2$ in tables of integrated values.

Except for the total flux density, the color bars in Figure \ref{fig:grmhd_hi_tables} reflect the $1\sigma$ values used to make cuts when evaluating models in \citetalias{PaperVIII}.

\begin{figure*}
    \centering
    \includegraphics[width=0.85\textwidth]{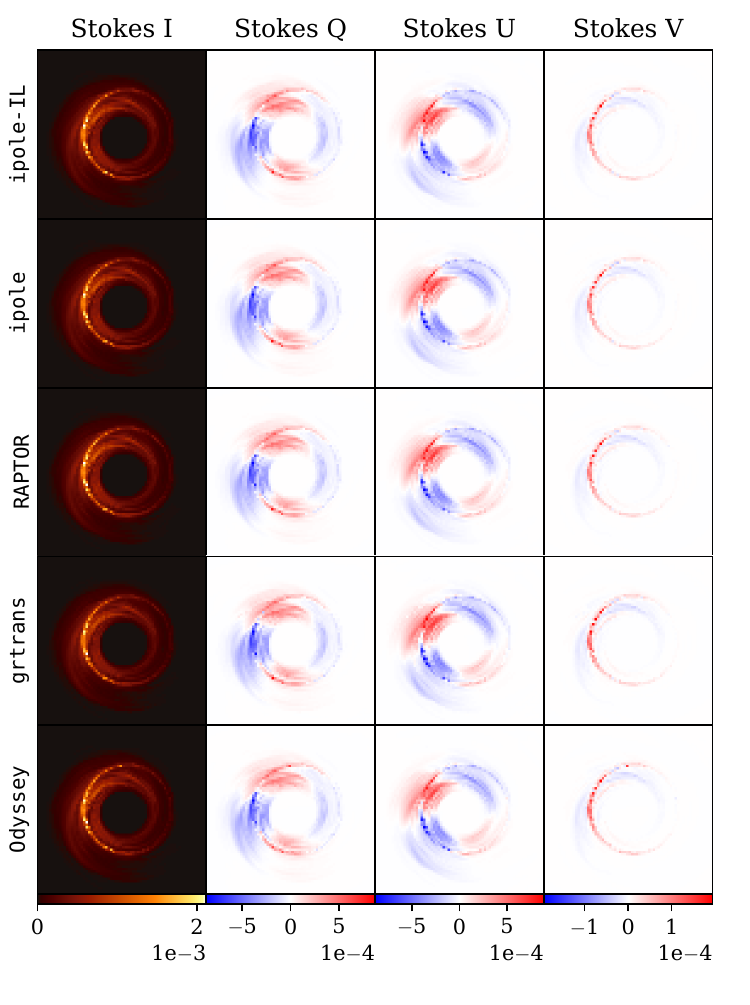}
    \caption{Comparison of images produced of the GRMHD test problem, in which codes produced an image given a simulated fluid snapshot, mirroring the analysis pipeline used in e.g. \citetalias{PaperV,PaperVIII}.  The Stokes parameters $I$, $Q$, $U$, $V$ are plotted separately, with separate color bars.}
    \label{fig:grmhd_hi_result}
\end{figure*}

\begin{deluxetable*}{l|ccccccc}
    \tablehead{ \colhead{Code } & \colhead{Flux [Jy]} & \colhead{$\mnet$ [\%]} & \colhead{$\vnet$ [\%]} & \colhead{$\mavg$ [\%]} & \colhead{EVPA [$^\circ$]} & \colhead{$\left| \beta_2 \right|$} & \colhead{$\angle \beta_2$ [$^\circ$]}}
    \startdata
\texttt{ipole-IL} & 0.47976 & 1.5232 & 0.64637 & 31.264 & -78.241 & 0.28165 & -20.307 \\
\texttt{ipole} & 0.47335 & 1.5904 & 0.62402 & 31.27 & -77.243 & 0.28171 & -20.622 \\
\texttt{RAPTOR} & 0.49396 & 1.6057 & 0.67381 & 31.893 & -77.797 & 0.28346 & -21.31 \\
\texttt{grtrans} & 0.45346 & 1.5135 & 0.69909 & 31.197 & -77.647 & 0.28063 & -17.903 \\
\texttt{Odyssey} & 0.46671 & 1.3928 & 0.70068 & 31.54 & -71.432 & 0.28325 & -20.713
    \enddata
    \caption{Image-integrated values for the GRMHD snapshot test.  Definitions for all values are given in Section \ref{sec:metrics}. Note that the EVPA is not used as a comparison metric in Figure \ref{fig:grmhd_hi_tables}.}
    \label{tab:grmhd_hi_int}
\end{deluxetable*}

\section{Discussion}\label{sec:discussion}

\subsection{Comparison to Observational Constraints}

The values for each comparison metric used as cuts in \citetalias{PaperVIII} are listed in Table \ref{tab:paper8_ranges}.  The table values are based on $1\sigma$ ranges for measurements of the same quantities in EHT data, described in \citetalias{PaperVII}.  These ranges provide a comparison to evaluate code interchangeability---if the differences between codes are substantially less than the range of measurement uncertainties, the analysis is agnostic to the choice of code employed.  The $1\sigma$ ranges listed are also used as the color bar ranges in the colored table listings in Figure \ref{fig:grmhd_hi_tables}.

\begin{deluxetable}{c|ccc}
    \tablehead{ \colhead{Parameter} & 
    \colhead{$1\sigma$} & \colhead{Code Uncertainty} & \colhead{Uncertainty/$\sigma$}}
    \startdata
    $\mnet$ 
    & 1.35 \% & 0.21 \% & 0.16 \\
    $\vnet$ 
    & 0.4 \% & 0.08 \% & 0.20 \\
    $\mavg$ 
    & 2.5 \% & 0.70 \% & 0.28 \\
    $\btwo$ 
    & 0.015 & 0.0026 & 0.17\\
    $\btwoangle$ 
    & 17$^\circ$ & 3.4$^\circ$ & 0.20
    \enddata
    \caption{The ``$1\sigma$'' column lists the $1\sigma$ detector uncertainty of each parameter, as estimated in \citetalias{PaperVII} and used as an allowable range when comparing models in \citetalias{PaperVIII}.  Note that the measurement of $\mavg$ was an upper bound; this upper bound was doubled to select the cut value. The ``Code Uncertainty'' column lists the greatest observed difference between codes when computing each parameter, and the final column lists this value as a proportion of the $1\sigma$ value.}
    \label{tab:paper8_ranges}
\end{deluxetable}

This comparison provides evidence that model evaluations as in \citetalias{PaperVIII} remain similar regardless of which of the included codes is employed.   As recorded in Table \ref{tab:paper8_ranges}, maximum code variation is universally less than 30\% of the detector $1\sigma$ range: in $\mnet$ (0.21\% vs 1.5\% ), $\vnet$ (0.08\% vs 0.4\%), $\mavg$ (0.70\% vs 2.5\%), $\btwo$ (0.0026 vs 0.015), and $\btwoangle$ (3.4$^\circ$ vs 17$^\circ$).  In any analysis based on cuts, code differences can shift a few particular images into or out of the final consideration. However, at these uncertainties no image from outside the $1\sigma$ detector uncertainty would be consistent with the central observed value.

Broadening the comparison to different images and models shows promising similarities. Appendix \ref{app:full_run} presents distributions of the image differences between \ipoleIL and \grtrans when run over thousands of snapshots of a very different model from the example: they show a wide variance but a smaller difference on average than in the example image, suggesting that image differences, at least between these codes, are mostly stochastic, further suppressing any potential effect on a cuts-based analysis as in \citetalias{PaperVIII}.

\subsection{Potential Measurement of System Parameters}
\label{sec:mse_constraints}

To translate the NMSE into a measure of code accuracy in testing model parameters, we define an ``error budget'' for each Stokes parameter, consisting of the largest NMSE between code results: 0.02 in $I$, 0.04 in $Q$ and $U$, and 0.13 in $V$.  Assuming perfect detector accuracy and modeling, this error characterizes which images are too similar to be effectively distinguished above code-to-code variations.

\begin{figure*}
    \includegraphics[width=\textwidth]{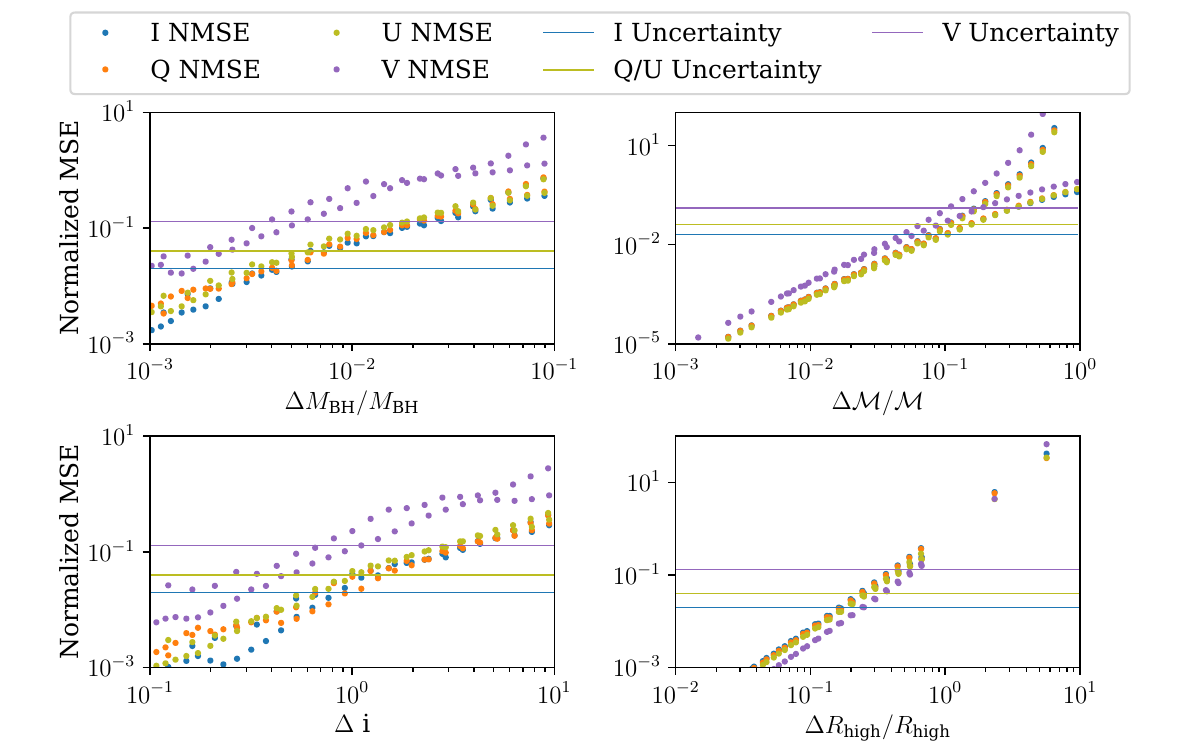}
    \caption{Plots of image mean squared error produced when varying several parameters (BH mass, angle, $\rhigh$, and accretion rate) around values used for the GRMHD snapshot test.  For each difference, plotted along the $x$ axis, images were run with the parameter increased and decreased by that amount, resulting in two NMSE values, which are then overplotted.  The largest NMSE between codes is plotted to provide a visual ceiling on ``indistinguishable'' images.  Based on code agreement, current transport schemes can constrain all input parameters much more accurately than current models and detectors.}
    \label{fig:MSE_vs_inputs}
\end{figure*}

We then translate this error budget into constraints on the input parameters $M_\mathrm{BH}$, $\mathcal{M}$, $\rhigh$, and the viewing angle by varying these parameters around the nominal values and calculating the resulting MSE vs.~the nominal image, using \ipoleIL.  As illustrated in Figure \ref{fig:MSE_vs_inputs}, the required parameter changes are very modest; that is, the possible constraints on system parameters are very precise.  In imaging the example model, codes agree well enough to constrain the mass of \bhname to within $0.4\%$ ($2.5 \times 10^{7} M_{\odot}$), $\mathcal{M}$ to $9\%$ ($1.5 \times 10^{25}$), the observer angle to $0.8^{\circ}$, and the $\rhigh$ parameter to within $0.18$. These values are dependent on the base image---in particular, constraints on $\rhigh$ will also depend on $\rlow$, which was set differently in this case than for images in \citetalias{PaperVIII}.

\subsection{Caveats and limitations}\label{sec:caveats}

There are a few caveats and limitations of the ray-tracing calculations presented in this work worth mentioning and improving in the future. Most glaringly, all ray-tracing codes use phenomenological post-processing models of the electron energy distribution.  In particular, this comparison adopts a fixed ratio of ion to electron temperature, which is not well-motivated by EHT results. More accurate temperature prescriptions including cold electrons in the accretion disk dramatically increase Faraday rotation when viewed from the equator, scrambling the emission angle over regions of the image.  Scrambled emission does not affect the total intensity image, nor the measurable quantities in this comparison (except $\btwoangle$, which only makes sense to measure for face-on images with low Faraday scrambling).  Detailed study of code behavior in imaging Faraday-scrambled regions is left for future work.

All ray-tracing codes use synchrotron emissivities/absorptivities/rotativities in analytic forms which are fit formulas to synchrotron emissions integrated over (most often thermal) electron distribution function. The fit functions may differ from code to code and from the true emissivity and therefore introduce a small error to the integration. We discuss this issue in more detail in Appendix~\ref{app:coefficients}. Other caveats concern the common assumption that the electrons are distributed isotropically, which may not be a good approximation for collisionless plasma surrounding Sgr~A* and M87*.

Calculations presented in this work assume the infinite speed of light (so called fast-light approximation) while in reality the light propagation timescale is comparable to the plasma dynamical timescale near the even horizon of the black hole. Any future comparison of ray-tracing codes should include finite light propagation time effects. In such future comparison another source of error could be the time-interpolations between GRMHD model time slices.

Finally, the linear polarization and the EVPA are  sensitive to the external Faraday screen made of mildly or non-relativistic electrons (which in practice could be located thousands of M away from the black hole). Any inconsistencies in choosing the outer boundary of the ray-tracing integration may introduce discrepancies in the linear polarization maps. The latter is not specific caveat of the ray-tracing itself but it is a limitation when comparing models to observations. 

\section{Conclusions}\label{sec:conclusions}

In each of the tests conducted for this comparison, the several general-relativistic radiative transfer codes used within the EHT Collaboration have produced sufficiently similar results that they are functionally interchangeable for the collaboration's uses.  This is true both when measured in terms of image similarity (mean squared error) and when measured directly in terms of the image metrics used to compare simulated polarized images to the EHT result in \citetalias{PaperVIII}.

Using their default accuracy parameters, codes match the analytic result for the case of constant transport coefficients to better than 1 part in $10^{-5}$.  They agree to within 1.5\% mean squared error when imaging an analytically-defined problem requiring parallel transport of the polarization vector.

In the more complex task of interpolating, translating, and imaging GRMHD output, codes agree to within a normalized mean squared error of 0.13 at worst, when measuring specifically the circular polarization map (NMSE of 0.045 in linear polarization, 0.02 in total intensity).  Based on image similarity, the choice of imaging code will matter in model comparisons only when trying to determine the BH mass to within $0.4\%$, accretion rate within $9\%$, or observer angle to within $0.8^{\circ}$. These values significantly outclass both the detection and modeling uncertainties available in the near future. 

When measured with the image metrics used for model comparison in \citetalias{PaperVIII}, all comparison images agree to much better precision than the detector uncertainty.  Further, much of the difference which does appear is shown to be stochastic in nature.  Thus the choice of code is verified directly to have little effect on the analysis performed in that paper.

\section{Acknowledgements}

The Event Horizon Telescope Collaboration thanks the following
organizations and programs: the Academia Sinica; the Academy
of Finland (projects 274477, 284495, 312496, 315721); the Agencia Nacional de Investigaci\'{o}n 
y Desarrollo (ANID), Chile via NCN$19\_058$ (TITANs) and Fondecyt 1221421, the Alexander
von Humboldt Stiftung; an Alfred P. Sloan Research Fellowship;
Allegro, the European ALMA Regional Centre node in the Netherlands, the NL astronomy
research network NOVA and the astronomy institutes of the University of Amsterdam, Leiden University and Radboud University;
the ALMA North America Development Fund; the Black Hole Initiative, which is funded by grants from the John Templeton Foundation and the Gordon and Betty Moore Foundation (although the opinions expressed in this work are those of the author(s) 
and do not necessarily reflect the views of these Foundations); the Brinson Foundation; 
Chandra DD7-18089X and TM6-17006X; the China Scholarship
Council; the China Postdoctoral Science Foundation fellowships (2020M671266, 2022M712084); Consejo Nacional de Ciencia y Tecnolog\'{\i}a (CONACYT,
Mexico, projects  U0004-246083, U0004-259839, F0003-272050, M0037-279006, F0003-281692,
104497, 275201, 263356);
the Consejer\'{i}a de Econom\'{i}a, Conocimiento, 
Empresas y Universidad 
of the Junta de Andaluc\'{i}a (grant P18-FR-1769), the Consejo Superior de Investigaciones 
Cient\'{i}ficas (grant 2019AEP112);
the Delaney Family via the Delaney Family John A.
Wheeler Chair at Perimeter Institute; Direcci\'{o}n General
de Asuntos del Personal Acad\'{e}mico-Universidad
Nacional Aut\'{o}noma de M\'{e}xico (DGAPA-UNAM,
projects IN112417 and IN112820); 
the Dutch Organization for Scientific Research (NWO) for VICI award (grant 639.043.513), grant OCENW.KLEIN.113 and the Dutch Black Hole Consortium (with project number NWA 1292.19.202) of the research programme the National Science Agenda; the Dutch National Supercomputers, Cartesius and Snellius  
(NWO Grant 2021.013); 
the EACOA Fellowship awarded by the East Asia Core
Observatories Association, which consists of the Academia Sinica Institute of Astronomy and
Astrophysics, the National Astronomical Observatory of Japan, Center for Astronomical Mega-Science,
Chinese Academy of Sciences, and the Korea Astronomy and Space Science Institute; 
the European Research Council (ERC) Synergy
Grant ``BlackHoleCam: Imaging the Event Horizon
of Black Holes" (grant 610058); 
the European Union Horizon 2020
research and innovation programme under grant agreements
RadioNet (No 730562) and 
M2FINDERS (No 101018682); the Horizon ERC Grants 2021 programme under grant agreement No. 101040021;
the Generalitat
Valenciana postdoctoral grant APOSTD/2018/177 and
GenT Program (project CIDEGENT/2018/021); MICINN Research Project PID2019-108995GB-C22;
the European Research Council for advanced grant `JETSET: Launching, propagation and 
emission of relativistic jets from binary mergers and across mass scales' (Grant No. 884631); 
the Institute for Advanced Study; the Istituto Nazionale di Fisica
Nucleare (INFN) sezione di Napoli, iniziative specifiche
TEONGRAV; 
the International Max Planck Research
School for Astronomy and Astrophysics at the
Universities of Bonn and Cologne; 
DFG research grant ``Jet physics on horizon scales and beyond'' (Grant No. FR 4069/2-1);
Joint Columbia/Flatiron Postdoctoral Fellowship, 
research at the Flatiron Institute is supported by the Simons Foundation; 
the Japan Ministry of Education, Culture, Sports, Science and Technology (MEXT; grant JPMXP1020200109); 
the Japan Society for the Promotion of Science (JSPS) Grant-in-Aid for JSPS
Research Fellowship (JP17J08829); the Joint Institute for Computational Fundamental Science, Japan; the Key Research
Program of Frontier Sciences, Chinese Academy of
Sciences (CAS, grants QYZDJ-SSW-SLH057, QYZDJSSW-SYS008, ZDBS-LY-SLH011); 
the Leverhulme Trust Early Career Research
Fellowship; the Max-Planck-Gesellschaft (MPG);
the Max Planck Partner Group of the MPG and the
CAS; the MEXT/JSPS KAKENHI (grants 18KK0090, JP21H01137,
JP18H03721, JP18K13594, 18K03709, JP19K14761, 18H01245, 25120007); the Malaysian Fundamental Research Grant Scheme (FRGS) FRGS/1/2019/STG02/UM/02/6; the MIT International Science
and Technology Initiatives (MISTI) Funds; 
the Ministry of Science and Technology (MOST) of Taiwan (103-2119-M-001-010-MY2, 105-2112-M-001-025-MY3, 105-2119-M-001-042, 106-2112-M-001-011, 106-2119-M-001-013, 106-2119-M-001-027, 106-2923-M-001-005, 107-2119-M-001-017, 107-2119-M-001-020, 107-2119-M-001-041, 107-2119-M-110-005, 107-2923-M-001-009, 108-2112-M-001-048, 108-2112-M-001-051, 108-2923-M-001-002, 109-2112-M-001-025, 109-2124-M-001-005, 109-2923-M-001-001, 110-2112-M-003-007-MY2, 110-2112-M-001-033, 110-2124-M-001-007, and 110-2923-M-001-001);
the Ministry of Education (MoE) of Taiwan Yushan Young Scholar Program;
the Physics Division, National Center for Theoretical Sciences of Taiwan;
the National Aeronautics and
Space Administration (NASA, Fermi Guest Investigator
grant 80NSSC20K1567, NASA Astrophysics Theory Program grant 80NSSC20K0527, NASA NuSTAR award 
80NSSC20K0645); 
NASA Hubble Fellowship 
grants HST-HF2-51431.001-A, HST-HF2-51482.001-A awarded 
by the Space Telescope Science Institute, which is operated by the Association of Universities for 
Research in Astronomy, Inc., for NASA, under contract NAS5-26555; 
the National Institute of Natural Sciences (NINS) of Japan; the National
Key Research and Development Program of China
(grant 2016YFA0400704, 2017YFA0402703, 2016YFA0400702); the National
Science Foundation (NSF, grants AST-0096454,
AST-0352953, AST-0521233, AST-0705062, AST-0905844, AST-0922984, AST-1126433, AST-1140030,
DGE-1144085, AST-1207704, AST-1207730, AST-1207752, MRI-1228509, OPP-1248097, AST-1310896, AST-1440254, 
AST-1555365, AST-1614868, AST-1615796, AST-1715061, AST-1716327,  AST-1716536, OISE-1743747, AST-1816420, AST-1935980, AST-2034306); 
NSF Astronomy and Astrophysics Postdoctoral Fellowship (AST-1903847); 
the Natural Science Foundation of China (grants 11650110427, 10625314, 11721303, 11725312, 11873028, 11933007, 11991052, 11991053, 12192220, 12192223); 
the Natural Sciences and Engineering Research Council of
Canada (NSERC, including a Discovery Grant and
the NSERC Alexander Graham Bell Canada Graduate
Scholarships-Doctoral Program); the National Youth
Thousand Talents Program of China; the National Research
Foundation of Korea (the Global PhD Fellowship
Grant: grants NRF-2015H1A2A1033752, the Korea Research Fellowship Program:
NRF-2015H1D3A1066561, Brain Pool Program: 2019H1D3A1A01102564, 
Basic Research Support Grant 2019R1F1A1059721, 2021R1A6A3A01086420, 2022R1C1C1005255); 
Netherlands Research School for Astronomy (NOVA) Virtual Institute of Accretion (VIA) postdoctoral fellowships; 
Onsala Space Observatory (OSO) national infrastructure, for the provisioning
of its facilities/observational support (OSO receives
funding through the Swedish Research Council under
grant 2017-00648);  the Perimeter Institute for Theoretical
Physics (research at Perimeter Institute is supported
by the Government of Canada through the Department
of Innovation, Science and Economic Development
and by the Province of Ontario through the
Ministry of Research, Innovation and Science); the Princeton Gravity Initiative; the Spanish Ministerio de Ciencia e Innovaci\'{o}n (grants PGC2018-098915-B-C21, AYA2016-80889-P,
PID2019-108995GB-C21, PID2020-117404GB-C21); 
the University of Pretoria for financial aid in the provision of the new 
Cluster Server nodes and SuperMicro (USA) for a SEEDING GRANT approved towards these 
nodes in 2020;
the Shanghai Pilot Program for Basic Research, Chinese Academy of Science, 
Shanghai Branch (JCYJ-SHFY-2021-013);
the State Agency for Research of the Spanish MCIU through
the ``Center of Excellence Severo Ochoa'' award for
the Instituto de Astrof\'{i}sica de Andaluc\'{i}a (SEV-2017-
0709); the Spinoza Prize SPI 78-409; the South African Research Chairs Initiative, through the 
South African Radio Astronomy Observatory (SARAO, grant ID 77948),  which is a facility of the National 
Research Foundation (NRF), an agency of the Department of Science and Innovation (DSI) of South Africa; 
the Toray Science Foundation; the Swedish Research Council (VR); 
the US Department
of Energy (USDOE) through the Los Alamos National
Laboratory (operated by Triad National Security,
LLC, for the National Nuclear Security Administration
of the USDOE (Contract 89233218CNA000001); and the YCAA Prize Postdoctoral Fellowship.

We thank
the staff at the participating observatories, correlation
centers, and institutions for their enthusiastic support.
This paper makes use of the following ALMA data:
ADS/JAO.ALMA\#2016.1.01154.V. ALMA is a partnership
of the European Southern Observatory (ESO;
Europe, representing its member states), NSF, and
National Institutes of Natural Sciences of Japan, together
with National Research Council (Canada), Ministry
of Science and Technology (MOST; Taiwan),
Academia Sinica Institute of Astronomy and Astrophysics
(ASIAA; Taiwan), and Korea Astronomy and
Space Science Institute (KASI; Republic of Korea), in
cooperation with the Republic of Chile. The Joint
ALMA Observatory is operated by ESO, Associated
Universities, Inc. (AUI)/NRAO, and the National Astronomical
Observatory of Japan (NAOJ). The NRAO
is a facility of the NSF operated under cooperative agreement
by AUI.
This research used resources of the Oak Ridge Leadership Computing Facility at the Oak Ridge National
Laboratory, which is supported by the Office of Science of the U.S. Department of Energy under Contract
No. DE-AC05-00OR22725. We also thank the Center for Computational Astrophysics, National Astronomical Observatory of Japan.
The computing cluster of Shanghai VLBI correlator supported by the Special Fund 
for Astronomy from the Ministry of Finance in China is acknowledged.
This work was supported by FAPESP (Fundacao de Amparo a Pesquisa do Estado de Sao Paulo) under grant 2021/01183-8.

APEX is a collaboration between the
Max-Planck-Institut f{\"u}r Radioastronomie (Germany),
ESO, and the Onsala Space Observatory (Sweden). The
SMA is a joint project between the SAO and ASIAA
and is funded by the Smithsonian Institution and the
Academia Sinica. The JCMT is operated by the East
Asian Observatory on behalf of the NAOJ, ASIAA, and
KASI, as well as the Ministry of Finance of China, Chinese
Academy of Sciences, and the National Key Research and Development
Program (No. 2017YFA0402700) of China
and Natural Science Foundation of China grant 11873028.
Additional funding support for the JCMT is provided by the Science
and Technologies Facility Council (UK) and participating
universities in the UK and Canada. 
The LMT is a project operated by the Instituto Nacional
de Astr\'{o}fisica, \'{O}ptica, y Electr\'{o}nica (Mexico) and the
University of Massachusetts at Amherst (USA). The
IRAM 30-m telescope on Pico Veleta, Spain is operated
by IRAM and supported by CNRS (Centre National de
la Recherche Scientifique, France), MPG (Max-Planck-Gesellschaft, Germany) 
and IGN (Instituto Geogr\'{a}fico
Nacional, Spain). The SMT is operated by the Arizona
Radio Observatory, a part of the Steward Observatory
of the University of Arizona, with financial support of
operations from the State of Arizona and financial support
for instrumentation development from the NSF.
Support for SPT participation in the EHT is provided by the National Science Foundation through award OPP-1852617 
to the University of Chicago. Partial support is also 
provided by the Kavli Institute of Cosmological Physics at the University of Chicago. The SPT hydrogen maser was 
provided on loan from the GLT, courtesy of ASIAA.

This work used the
Extreme Science and Engineering Discovery Environment
(XSEDE), supported by NSF grant ACI-1548562,
and CyVerse, supported by NSF grants DBI-0735191,
DBI-1265383, and DBI-1743442. XSEDE Stampede2 resource
at TACC was allocated through TG-AST170024
and TG-AST080026N. XSEDE JetStream resource at
PTI and TACC was allocated through AST170028.
This research is part of the Frontera computing project at the Texas Advanced 
Computing Center through the Frontera Large-Scale Community Partnerships allocation
AST20023. Frontera is made possible by National Science Foundation award OAC-1818253.
This research was done using services provided by the OSG Consortium~\citep{osg07,osg09}, which is supported by the National Science Foundation awards \#2030508 and \#1836650.
Additional work used ABACUS2.0, which is part of the eScience center at Southern Denmark University. 
Simulations were also performed on the SuperMUC cluster at the LRZ in Garching, 
on the LOEWE cluster in CSC in Frankfurt, on the HazelHen cluster at the HLRS in Stuttgart, 
and on the Pi2.0 and Siyuan Mark-I at Shanghai Jiao Tong University.
The computer resources of the Finnish IT Center for Science (CSC) and the Finnish Computing 
Competence Infrastructure (FCCI) project are acknowledged. This
research was enabled in part by support provided
by Compute Ontario (http://computeontario.ca), Calcul
Quebec (http://www.calculquebec.ca) and Compute
Canada (http://www.computecanada.ca). 

The EHTC has
received generous donations of FPGA chips from Xilinx
Inc., under the Xilinx University Program. The EHTC
has benefited from technology shared under open-source
license by the Collaboration for Astronomy Signal Processing
and Electronics Research (CASPER). The EHT
project is grateful to T4Science and Microsemi for their
assistance with Hydrogen Masers. This research has
made use of NASA's Astrophysics Data System. We
gratefully acknowledge the support provided by the extended
staff of the ALMA, both from the inception of
the ALMA Phasing Project through the observational
campaigns of 2017 and 2018. We would like to thank
A. Deller and W. Brisken for EHT-specific support with
the use of DiFX. We thank Martin Shepherd for the addition of extra features in the Difmap software 
that were used for the CLEAN imaging results presented in this paper.
We acknowledge the significance that
Maunakea, where the SMA and JCMT EHT stations
are located, has for the indigenous Hawaiian people.


\appendix
\twocolumngrid

\section{Comparison of polarized spectra from imaging and Monte Carlo radiative transfer methods}\label{app:sed}

\begin{figure}
    \centering
    \includegraphics[width=0.4\textwidth,trim={0 0 0 2cm},clip]{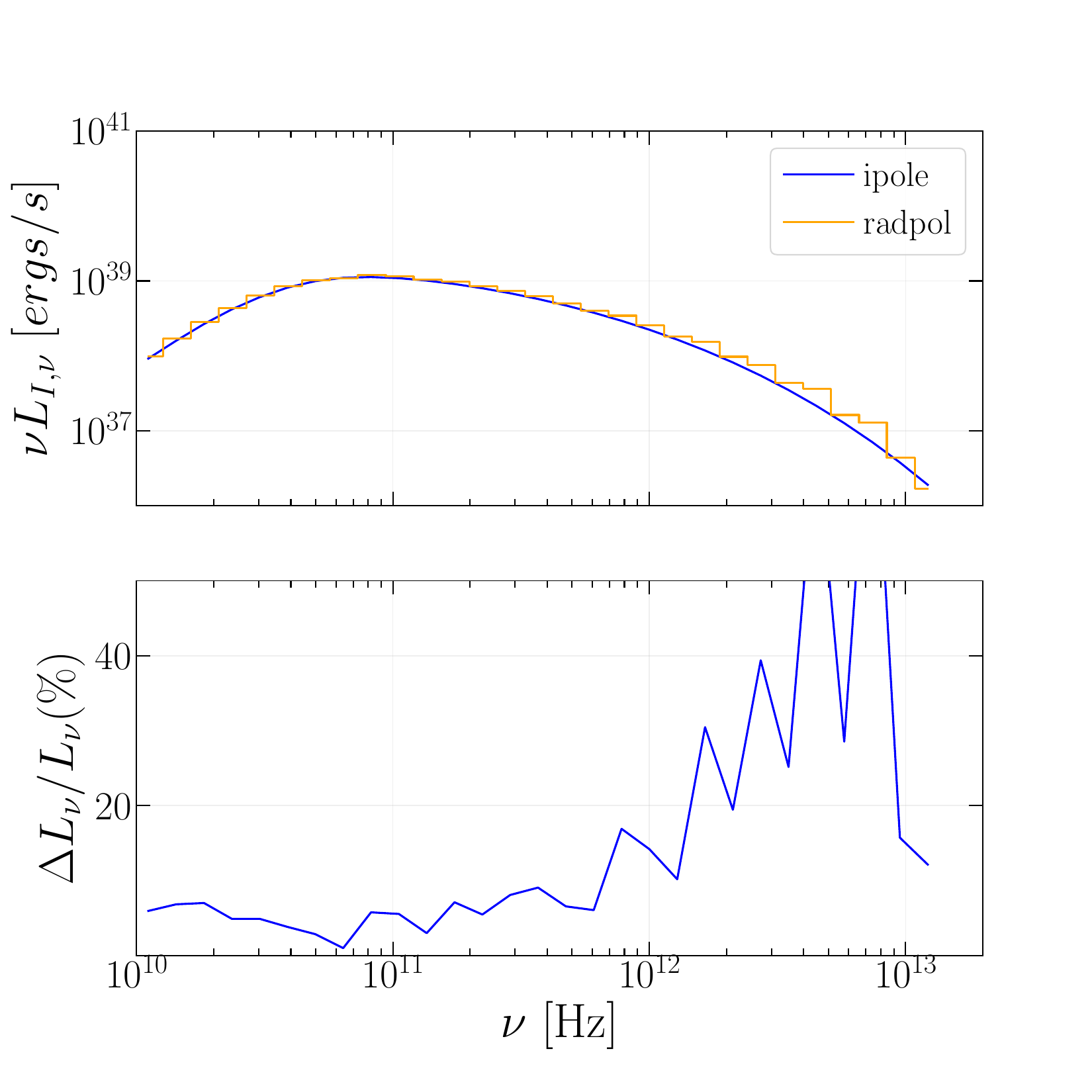}
    \includegraphics[width=0.4\textwidth,trim={0 0 0 2cm},clip]{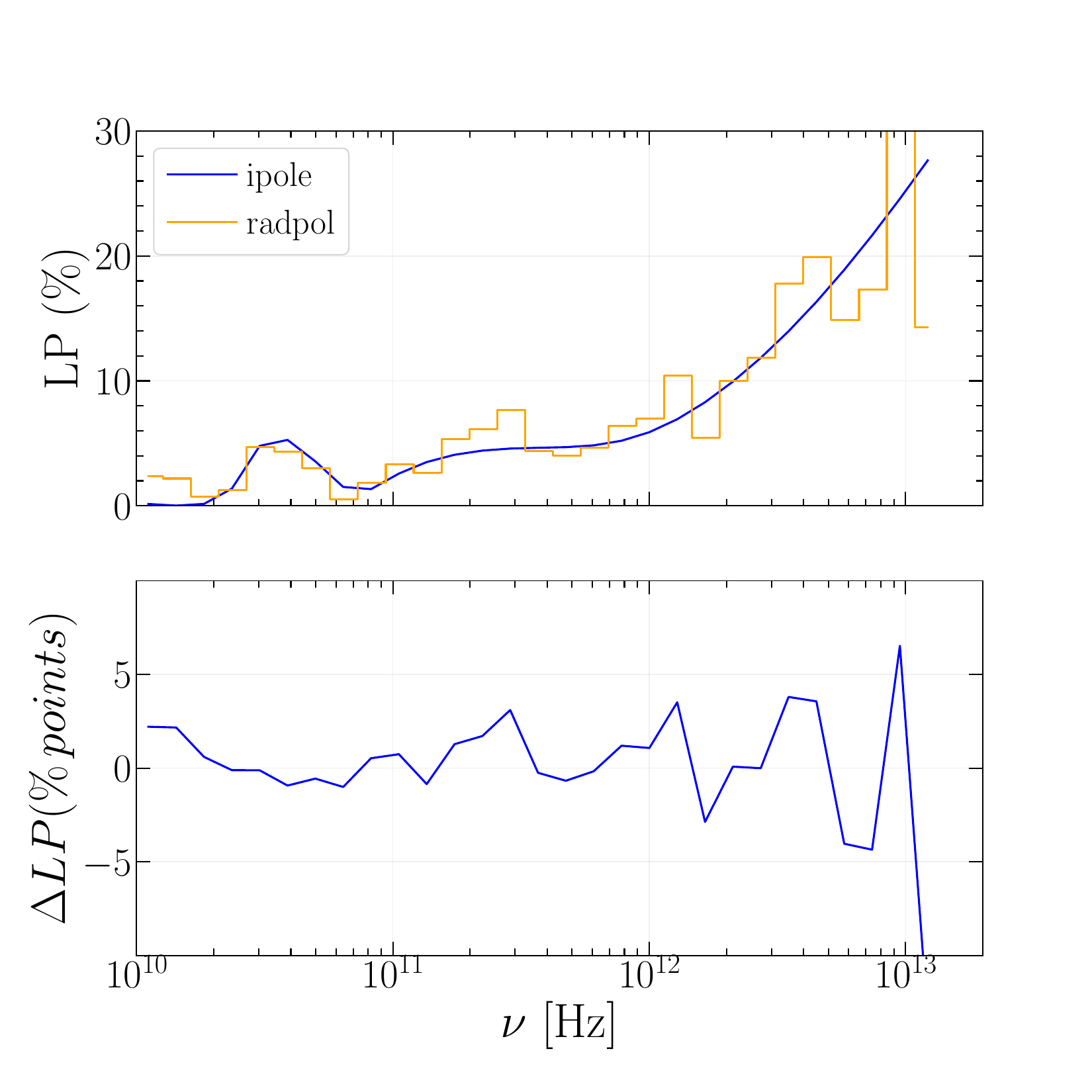}
    \includegraphics[width=0.4\textwidth,trim={0 0 0 2cm},clip]{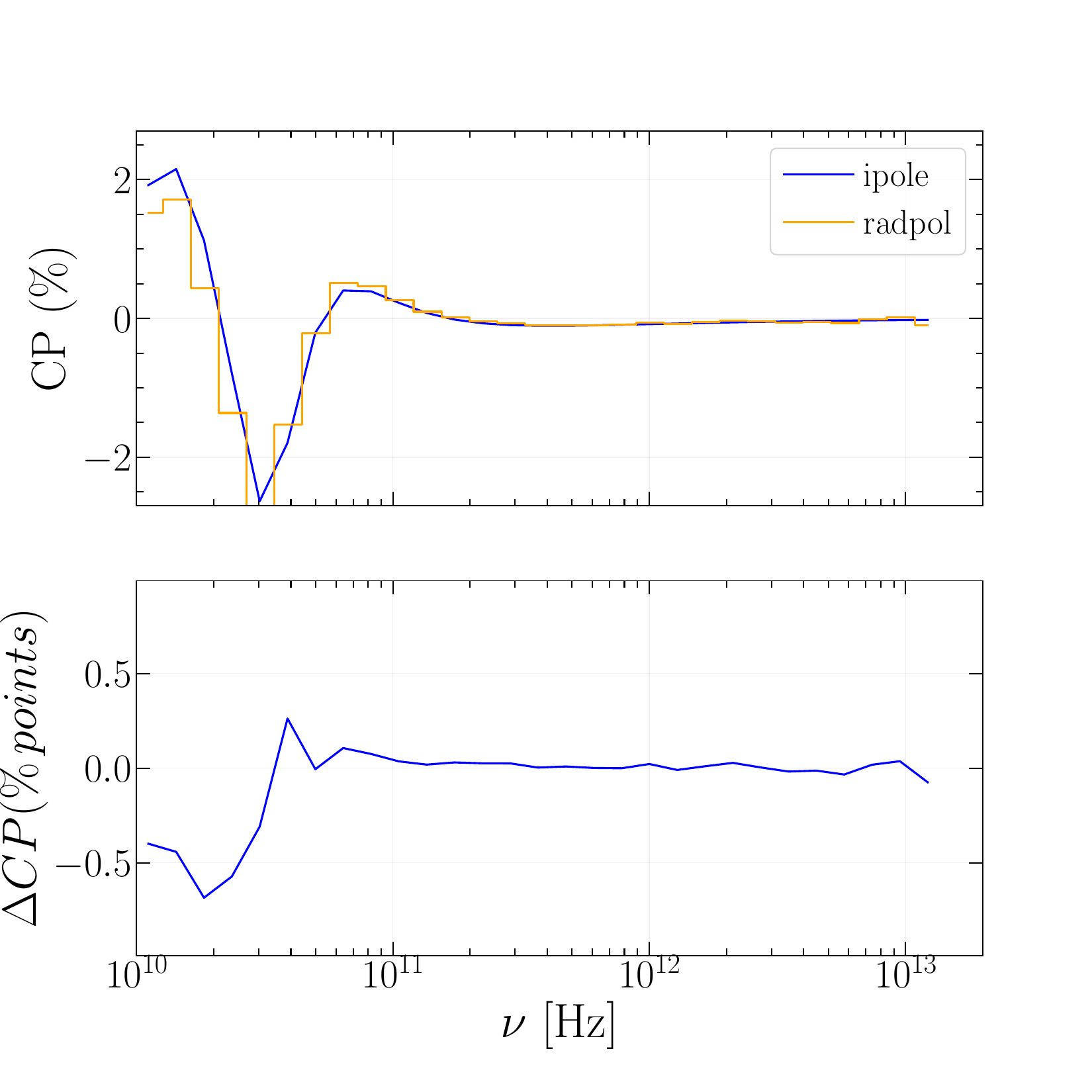}
    \caption{Comparison of luminosity, fractional linear and circular polarizations across synchrotron spectrum produced by {\tt ipole} and {\tt radpol} codes using SANE simulation with the same parameters.
    Both radiative transfer schemes show
consistent amplitude of fractional linear and circular polarizations and agree on handedness of circular polarization.}
    \label{fig:img_mc}
\end{figure}

In all imaging ray tracing codes radiative transfer equations are solved along null-geodesics that terminate at a ``camera'' at some large distance from the supermassive black holes and where a polarization map at a chosen observing frequency is constructed. By integrating Stokes parameters over entire images made for different frequencies one can also construct a polarized synchrotron spectral energy distribution of any model. However instead of comparing model spectra from the discussed imaging codes here we carry out an alternative comparison. Namely we compare spectra produced by {\tt ipole} code to polarized spectra generated via Monte Carlo scheme {\tt radpol}. In Monte Carlo code the polarized radiative transfer integration scheme is conceptually distinct from all discussed imaging codes (for detailed description see \citealt{Moscibrodzka2020}). Showing a convergence of two different approaches is an independent validation of emission produced by imaging codes. Hence, we compare spectra produced by {\tt ipole} and {\tt radpol} codes using plasma model setup described in Section~\ref{sec:prob_grmhd} (the test with $\mathcal{M}_{\rm low}$). In Figure~\ref{fig:img_mc}, we show radio-millimeter spectra of Stokes I luminosity, fractional linear polarization and circular polarizations. The relative difference between luminosities is less than 10\%, except for high frequency emission. Both codes show consistent amplitude of fractional linear and circular polarizations and agree on handedness of circular polarization.

\section{Effect of Polarized Emissivity and Rotativity Fits}
\label{app:coefficients}

The emission, absorption, and rotation coefficients of a fluid are well-determined for a particular distribution of electron energies. However, the integrations involved are numerically expensive; since the coefficients must be calculated at every step when integrating the radiative transfer equations, fitting functions have been developed to approximate the coefficients quickly.  A few sets of such fitting functions exist applicable to our regime; one outlined in \citet{Dexter2016}, the other in \citet{Pandya2016} and \citet{Marszewski2021}.

The differences between these functions at various points within a representative set of input parameters are given in their respective papers, but we wish to provide some intuition concerning the differences these functions make in practice, and consequently whether fitting accuracy might be a driving factor in code differences.

All of the images in this comparison were created using the coefficient fits from \citet{Dexter2016}.  Substituting the coefficient fits from \citet{Pandya2016} produces results more dissimilar than the disagreement between codes on three metrics: the net circular polarization at 0.13 points rather than 0.08 points, and average linear polarization fraction at 1.8 points rather than 0.7 points, and the $\btwo$ coefficient at 0.016 rather than 0.0026.  This is due to significant differences between the fits in computing the emission coefficient for circularly polarized light, $j_V$.

\begin{deluxetable*}{l|cccc|ccccc}
    \tablehead{ \colhead{Image run with...} & \colhead{NMSE $I$} & \colhead{NMSE $Q$} & \colhead{NMSE $U$} & \colhead{NMSE $V$}
                & \colhead{$\Delta \mnet$ [\%]} & \colhead{$\Delta \mavg$ [\%]} & \colhead{$\Delta \vnet$ [\%]} & \colhead{$\Delta \btwo$} & \colhead{$\Delta \btwoangle$ [$^\circ$]} }
    \startdata
...Pandya $j_S$ & 4.7e-05 & 0.0055 & 0.0059 & 0.0087 & 0.0246 & 1.78 & 0.127 & 0.0168 & 0.353 \\
...Dexter $\rho_V$ & 1.4e-09 & 2.6e-06 & 2.1e-06 & 6.8e-06 & -0.00746 & 0.0186 & -0.0031 & 0.000131 & 0.00225 \\
...approx. $K_n(x)$ & 3.9e-08 & 5.6e-05 & 2.5e-05 & 0.00014 & -0.000404 & 0.0239 & -0.00524 & 0.000123 & -0.00336
    \enddata
    \caption{Comparison of changes to the \ipoleIL result under different emission coefficient fitting functions. Each row lists a change made to the default emissivity values and the resulting differences between the new image and the one used in the comparison.  Substituting emissivities from \citet{Pandya2016} changes the result by more than the overall level of code agreement, whereas substituting the $rho_V$ fit from \citet{Dexter2016} changes almost nothing about the image.  Approximating the Bessel functions $K_n$ does not badly affect this image, but can be a substantial source of error in images with higher Faraday rotation.}
    \label{tab:coefficients}
\end{deluxetable*}

Unlike most emission coefficients, Faraday rotation coefficient $\rho_V$ does not go to zero with low temperature -- therefore the low-temperature behavior of fitting functions is important.  In particular, the expression from \cite{Dexter2016} for $\rho_V$ should not be used at low temperature $\Theta_{\mathrm{e}} < 1$, since it can produce a catastrophic cancellation not matching the desired limiting behavior of one of its quotients. Due to this instability, most codes either switch to the Shcherbakov fit at low temperature (e.g., \grtrans), or use the Shcherbakov fit exclusively (e.g., \ipoleIL).  As illustrated in Table \ref{tab:coefficients} these approaches produce nearly identical results, different by a mean squared error less than $10^{-5}$ in the worst case.

Also, both the expressions from \cite{Shcherbakov2008} and \cite{Dexter2016} involve Bessel functions, which are tempting to approximate by assuming emission is exclusive to the regime $\Theta_{\mathrm{e}} > 1$ in order to avoid unnecessary computation.  However, this approximation produces clearly incorrect limiting behavior for $\rho_{\mathrm{V}}$ at low temperature.  Thus the Faraday rotation is misapplied, producing too little rotation in the EVPA, or in some cases, rotation in the wrong direction.  In the sample image this is a minor effect due to an overall small Faraday rotation, but it can severely affect SANE disks seen from larger observer angles.

\section{Comparison Between Image Difference Metrics}
\label{app:metric_compare}

In addition to the mean squared error, several other metrics could be used to gauge image dissimilarity between codes.  Three additional metrics were evaluated in the context of this comparison: the normalized mean linear error (NMLE), structural dissimilarity (DSSIM), and inverse zero-normalized cross correlation (DZNCC).  These are defined as follows:

\begin{align}
    \mathrm{NMLE}(A, B) &= \frac{\sum_j|A_j-B_j|}{\sum_j|A_j|} \\
    \mathrm{SSIM}(A, B) &= \left( \frac{2 \mu_A \mu_B}{\mu_A^2 + \mu_B^2} \right) \nonumber \\
    &\times \left( \frac{ \frac{2}{N} \sum_j \left( A_j - \mu_A \right) \left( B_j - \mu_B \right)}{\sigma_A^2 + \sigma_B^2} \right) \\
    \mathrm{DSSIM}(A, B) &= \frac{1}{\left| \mathrm{SSIM}(A, B) \right|} - 1 \\
    \mathrm{ZNCC}(A, B) &= \frac{1}{N} \sum_j \frac{1}{\sigma_A \sigma_B} \left( A_j - \mu_A \right) \left( B_j - \mu_B \right) \\
    \mathrm{DZNCC}(A, B) &= \frac{1}{\left| \mathrm{ZNCC}(A, B) \right|} - 1
\end{align}
where $\mu_X$ is the average pixel value of an image and $\sigma_X$ is the standard deviation of the pixel values.

\begin{figure*}
    \centering
    \includegraphics[width=\textwidth]{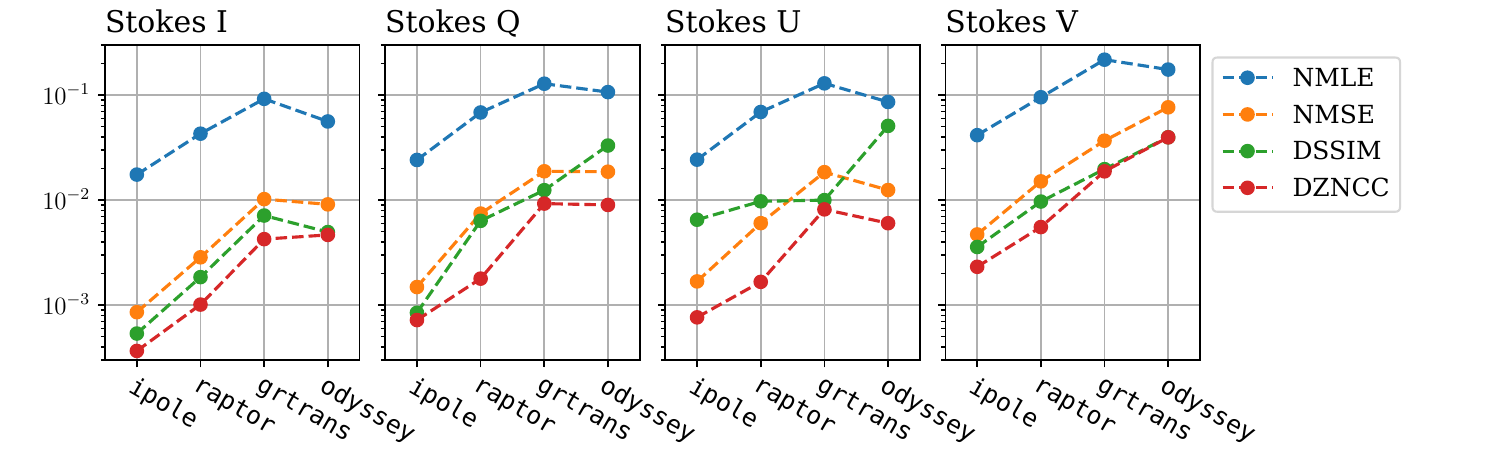}
    \caption{Each Stokes parameter of the \ipoleIL image result for the high-$\mathcal{M}$ GRMHD snapshot test, evaluated against each other code using 4 different metrics (all normalized): the mean linear error (MLE), mean squared error (MSE), structural dissimilarity (DSSIM), and inverse zero-normalized cross-correlation (DZNCC).}
    \label{fig:metric_diffs}
\end{figure*}

In this comparison, most images were very similar---for this limited case, the various similarity metrics were found to correlate strongly in ordering of similarity between all images, and usually even in relative magnitude, as shown in Figure \ref{fig:metric_diffs}.  The absolute values of the constraint metrics mean little without the context of Section \ref{sec:mse_constraints}, so any particular metric could fill the role of a similarity gauge to compare to image variation from other sources.

\section{Comparison of Images Over Full GRMHD Run}
\label{app:full_run}

While the GRMHD snapshot file used for the test in \ref{sec:prob_grmhd} reflects the simulations and imaging parameters used in practice by the EHTC in studies of \bhname, there is always the chance that the snapshot itself is a particularly simple case, not reflective of average code differences in practice.  Furthermore, in characterizing the impact of code differences on metric-based model comparisons, it would be useful to have an idea of what portion of code differences in metrics are due to systematic errors, vs.~stochastic products of limited accuracy parameters or sampling differences.

To measure the variation in results of this test over a typical variety of GRMHD states, two codes with substantially different algorithms, \ipoleIL and \grtrans, were compared across $2,000$ snapshots of a GRMHD simulation used in generating the EHT image libraries.  This particular simulation represented a magnetically arrested disk (MAD) state about a BH of spin $a_* = 0.9375$, and the $2,000$ snapshots shown represent the entire quiescent portion of the simulation from $5,000\;r_g/c$ to $15,000\;r_g/c$.  Details of the initial conditions, resolution, etc.~are available in \citet{Wong2022}.

\begin{figure*}
    \centering
    \includegraphics[width=\textwidth]{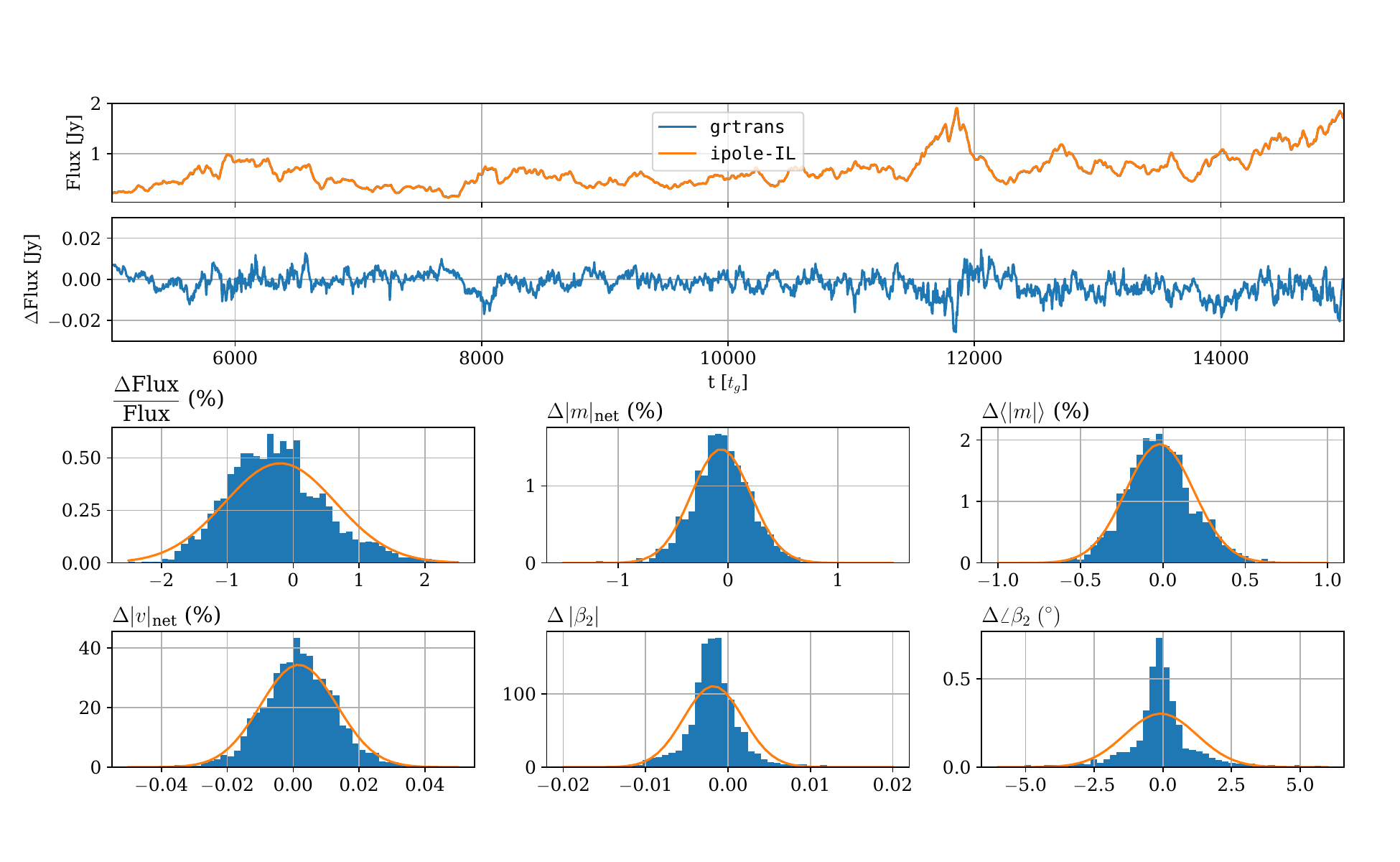}
    \caption{(upper plots) Comparison of total flux density computed by \ipoleIL and \grtrans over $10,000\;r_g/c$ of a MAD GRMHD simulation. The upper plot shows the total (Stokes I) flux density produced by each code, and the lower the absolute difference in flux densities.
    (lower plots) Histograms comparing image metrics between corresponding images over the entire window.  Each histogram is computed with a total of 50 bins across the domain shown, with any values outside the range added to the final bins.}
    \label{fig:fullrun_all}
\end{figure*}

Figure \ref{fig:fullrun_all} compares the total unpolarized flux computed by \ipoleIL and \grtrans over the entire window.  The lower plots provide histograms of the differences in all image-integrated values over the window, along with Gaussian functions following their means and standard deviations.  Table \ref{tab:fullrun_means} lists the mean (i.e.,~average difference) and standard deviation (i.e.,~span of differences) between codes in each metric.  The parameters $\mavg$ and $\btwoangle$ appear to be almost entirely stochastic. Flux, $\mnet$, and $\vnet$ are approximately a quarter systematic, and $\btwo$ is half systematic (though the error itself is minuscule).

\begin{deluxetable}{l|ccc}
    \vspace{0.5cm}
    \tablehead{ \colhead{Variable} & \colhead{$\mu$} & \colhead{$\sigma$} & \colhead{$\mu / \sigma$} }
    \startdata
$\frac{\Delta \mathrm{Flux}}{\mathrm{Flux}}$ (\%) & -0.197 & 0.841 & 0.234 \\
$\Delta |m|_\mathrm{net}$ (\%) & -0.0627 & 0.27 & 0.232 \\
$\Delta \langle |m| \rangle$ (\%) & -0.0182 & 0.206 & 0.0884 \\
$\Delta |v|_\mathrm{net}$ (\%) & 0.00168 & 0.0116 & 0.145 \\
$\Delta \left| \beta_2 \right|$ & -0.00178 & 0.00361 & 0.493 \\
$\Delta \angle \beta_2 \; (^\circ)$ & -0.0757 & 1.31 & 0.0576
    \enddata
    \caption{Mean values and standard deviations of the distributions in Figure \ref{fig:fullrun_all}.  The last column lists the proportion $\mu/\sigma$, which can be taken as an estimate of the relevance of systematic vs stochastic errors in describing differences between these codes.}
    \label{tab:fullrun_means}
    \vspace{-1cm}
\end{deluxetable}


\bibliographystyle{aasjournal}
\bibliography{pol_grrt_base,EHTC,osg}

\end{document}